\documentclass[10pt,journal,compsoc]{IEEEtran}
\ifCLASSOPTIONcompsoc
  \usepackage[nocompress]{cite}
\else
  \usepackage{cite}
\fi
\ifCLASSINFOpdf
\else
\fi

\usepackage{float}
\usepackage{graphicx}
\usepackage{amsmath}
\usepackage{amssymb}
\usepackage{multirow,hhline}
\usepackage[table]{xcolor}
\usepackage{booktabs}% for better rules in the table
\usepackage{url}
\usepackage{footnote}
\usepackage{hyperref}

\usepackage{array}
\newcommand{\PreserveBackslash}[1]{\let\temp=\\#1\let\\=\temp}
\newcolumntype{C}[1]{>{\PreserveBackslash\centering}p{#1}}
\newcolumntype{R}[1]{>{\PreserveBackslash\raggedleft}p{#1}}
\newcolumntype{L}[1]{>{\PreserveBackslash\raggedright}p{#1}}

\begin{document}
\title{A Little Bit More:\\Bitplane-Wise Bit-Depth Recovery}

\author{Abhijith~Punnappurath and Michael~S.~Brown, \IEEEmembership{Senior Member,~IEEE}% <-this % stops a space
\IEEEcompsocitemizethanks{\IEEEcompsocthanksitem A. Punnappurath and M. S. Brown are with Samsung AI Center, Toronto, Canada.
E-mail: abhijith.p@samsung.com, michael.b1@samsung.com
}% <-this % stops an unwanted space
\thanks{Manuscript received April 19, 2005; revised August 26, 2015.}}


\markboth{Journal of \LaTeX\ Class Files,~Vol.~14, No.~8, August~2015}%
{Shell \MakeLowercase{\textit{et al.}}: Bare Demo of IEEEtran.cls for Computer Society Journals}



\IEEEtitleabstractindextext{%
\begin{abstract}
Imaging sensors digitize incoming scene light at a dynamic range of 10--12 bits (i.e., 1024--4096 tonal values). The sensor image is then processed onboard the camera and finally quantized to only 8 bits (i.e., 256 tonal values) to conform to prevailing encoding standards. There are a number of important applications, such as high-bit-depth displays and photo editing, where it is beneficial to recover the lost bit depth. Deep neural networks are effective at this bit-depth reconstruction task. Given the quantized low-bit-depth image as input, existing deep learning methods employ a single-shot approach that attempts to either (1) directly estimate the high-bit-depth image, or (2) directly estimate the residual between the high- and low-bit-depth images. In contrast, we propose a training and inference strategy that recovers the residual image bitplane-by-bitplane. Our bitplane-wise learning framework has the advantage of allowing for multiple levels of supervision during training and is able to obtain state-of-the-art results using a simple network architecture. We test our proposed method extensively on several image datasets and demonstrate an improvement from 0.5dB to 2.3dB PSNR over prior methods depending on the quantization level.
\end{abstract}

}

\maketitle

\IEEEdisplaynontitleabstractindextext

% \fi
\IEEEpeerreviewmaketitle

\IEEEraisesectionheading{\section{Introduction}\label{sec:introduction}}

\IEEEPARstart{T}{he} term {\it bit depth} refers to the number of bits used per color channel to encode the tonal values of an image. Most modern camera sensors capture images with bit depths between 10 and 12 bits (representing 1024 and 4096 tonal values, respectively).  The captured sensor image is then processed onboard the camera to transform it to a standard RGB (sRGB) image that is compatible with prevailing encoding standards~\cite{hakki}.  The final step in the in-camera processing pipeline is to \emph{quantize} the sRGB image into an 8-bit-per-channel format to match the bit depth of consumer displays and printers, the vast majority of which are 8 bit.

\begin{figure}[!t]
%\begin{center}
\centering
\includegraphics[width=1.0\linewidth]{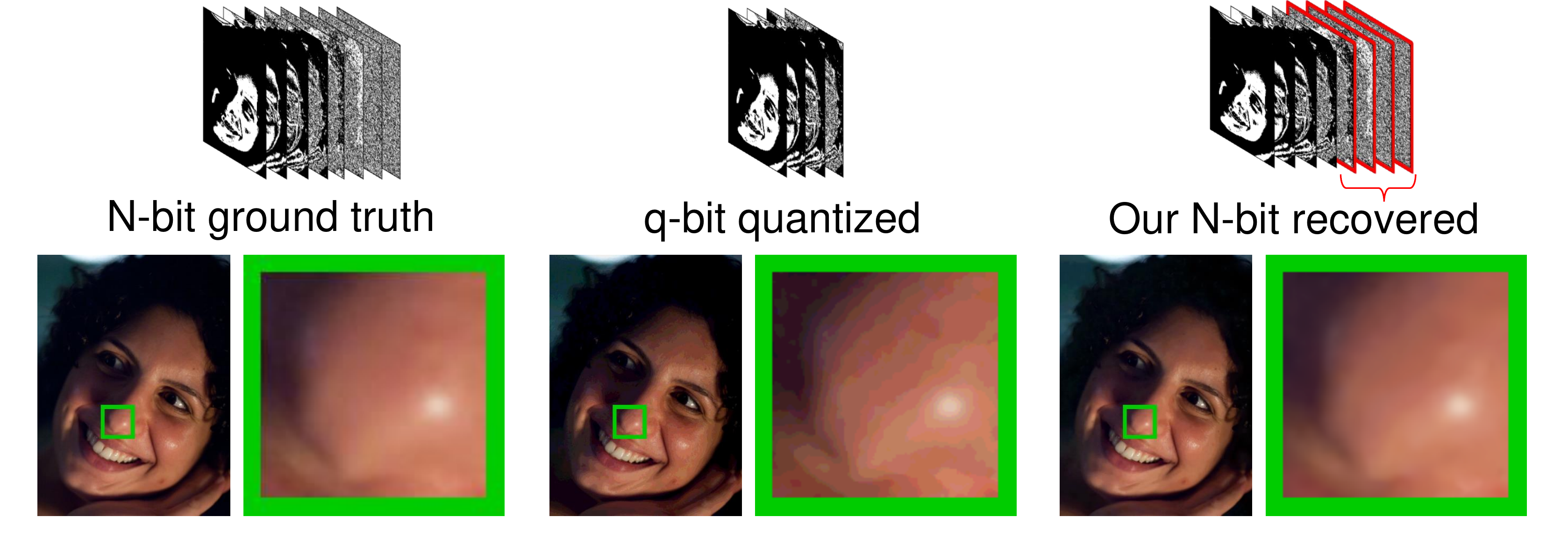}
%\end{center}
\caption{The ground truth $N$-bit high-bit-depth (HBD) image, its $q$-bit quantized low-bit-depth (LBD) version, and our recovered $N$-bit HBD output are shown. Our proposed bit-depth recovery algorithm restores the $q$-bit quantized input image to its original bit depth of $N$ by recovering the lost $(N$-$q)$ bits (shown in red), one bitplane at a time. The green box denotes a zoomed-in region of the image. While $(q,N)$=$(8,10)$ or $(8,12)$ are common target use cases in HBD displays or photo editing applications, most printers and monitors (and even this paper's PDF format) are designed for $8$-bit. Therefore, for the purpose of this example, we use $(q,N)$=$(4,8)$ so that the quantization effects can be visually observed in print and on screen.}
\label{fig:teaser}
\end{figure}

Bit-depth recovery is the task of recovering the bits lost due to quantization. An example is shown in Fig.~\ref{fig:teaser}. Bit-depth recovery has important applications in high-bit-depth (HBD) displays and photo editing.
%The ability to edit an image post-capture can be adversely affected by this quantization step. Fig. \ref{fig:teaser}-(A) shows a common example of a captured 8-bit sRGB image having low contrast. Applying a simple histogram equalization operation on this 8-bit image to enhance its contrast produces the undesired ``stair-step'' effect shown in Fig. \ref{fig:teaser}-(B).  This artifact is a direct consequence of the quantization process whereby tonal information was lost.  Fig. \ref{fig:teaser}-(B) shows the results of histogram equalization when applied to the original unquantized 12-bit image (which is usually unavailable) and our recovered 12-bit result. Due to the additional tonal values in the 12-bit histograms, the banding artifacts are not present in these outputs. This example highlights the need for bit-depth recovery.
While most monitors are still 8 bit, 10-bit display devices are becoming increasingly popular due to ever-growing consumer demand for finer and more nuanced color tones. Many TVs, and smart phones such as Samsung Galaxy S10 \cite{S10} and iPhone X \cite{iPhoneX}, already support 10-bit displays to meet high dynamic range (HDR) standards. However, these displays remain under-utilized as most image and video content available is still 8 bit.  Artifacts are observed if the bit depth is not restored before 8-bit data is displayed on 10-bit monitors.
%Fig.\ref{fig:applications} shows the outputs obtained without and with bit-depth recovery applied as a pre-processing step. Note that HBD displays may further process the input (e.g., apply different rendering styles) to enhance the subjective quality and produce visually pleasing results. While displays often apply additional image enhancement operations, bit-depth recovery is fundamentally an image restoration task whose objective is to recover the actual bits lost due to quantization.
Likewise, restoring the bit depth prior to photo editing reduces artifacts due to the availability of additional tonal values.
%, as shown in Fig.\ref{fig:applications}.

Several classical methods (e.g.,~\cite{BR,MRC,CRR,CA,ACDC,IPAD}) for bit-depth recovery exist in the literature. More recently, deep learning algorithms~\cite{BE-CNN,BE-CALF,BitNet,BDEN,Hou,GG-DCNN,jing:2018:bde} have been proposed. The strategy employed by most deep learning methods (e.g.,~\cite{BE-CNN,BitNet,BDEN,Hou,GG-DCNN,jing:2018:bde}) is to train the network using the quantized low-bit-depth (LBD) image as input and have the network predict the high-bit-depth image as output. A more recent alternative strategy~\cite{BE-CALF} is to train the network to predict the residual between the ground truth HBD image and the quantized LBD image. At test time, this output residual image is added to the quantized LBD image to produce the desired HBD image. While these two strategies are equivalent, it is argued that from a deep neural network (DNN) training perspective, the latter is more efficient and provides faster convergence. We adopt this latter formulation in our work. However, as opposed to~\cite{BE-CALF}, who estimate the residual in a single shot, we propose to recover the residual image corresponding to the lost bits one bitplane at a time.

Because the numerical magnitude of a pixel's tonal value decreases from the most significant bit to the least significant bit, the most significant bits dominate loss functions based on mean squared error (MSE) or peak signal-to-noise ratio (PSNR). As a result, approaches that predict either the HBD image or the residual using single-shot training fail to capture the very fine details encoded by the lower-order bits.  This observation is the impetus for our bitplane-wise training strategy that aims to overcome this limitation of existing methods.

\noindent \textbf{Contributions}~
Our proposed method is based on the observation that the residual between the high-bit-depth ground truth image and the quantized low-bit-depth input image can be expressed as a weighted summation of the bitplanes lost during quantization. Each bitplane is a binary map, and thus, the original task of predicting the residual can be reformulated into a series of better-constrained binary image segmentation problems. We train a separate neural network independently to predict each bitplane. This makes our method agnostic to the relative magnitude of the bit position, and overcome the limitation of the single-shot training strategy employed by existing approaches. The input images for training each network can be generated from the ground truth HBD image by applying the appropriate level of quantization. This multi-level-supervised training strategy outperforms state-of-the-art techniques using a simple network architecture. We compare our method against several classical as well as deep learning algorithms on five image datasets under a range of quantization levels and demonstrate that our method advances the state-of-the-art by a significant margin.
% THIS FIGURE HAS BEEN REMOVED IN THE REVISION
%\begin{figure}[!t]
%\begin{center}
%\includegraphics[width=1.0\linewidth]{figures/teaser6.pdf}
%\end{center}
%\caption{An illustrative example demonstrating the importance of bit-depth recovery applied as a pre-processing step for HBD displays and photo editing applications. Stretching the quantized LBD image for display on an HBD device produces artifacts as can be seen from row one, whereas our recovered HBD image when provided as input leads to high-quality results as shown in row two. Photo editors too derive similar advantages from the availability of additional tonal values. Bit-depth recovery is a restoration task, and provides a canonical starting point for enhancement operations of choice applied during HBD display rendering or photo editing.}
%\label{fig:applications}
%\end{figure}

\noindent \textbf{Scope}~
Our method is focused strictly on bit-depth quantization and does not consider other strategies often used in bit-depth reduction, such as image halftoning~\cite{sharma:2002:digital}, or color palette reduction, such as in GIF~\cite{GIF}.  In addition, our method is not attempting inverse-tone mapping~\cite{eilertsen:2017:hdr,Kim2019DeepSJ}, which is an image enhancement technique targeted at producing plausible high-dynamic-range images for HBD display from low-dynamic-range input data. In contrast, the aim of bit-depth recovery, and our algorithm in particular, is image restoration where the goal is to accurately recover the actual bits lost due to linear quantization.

% Original version

%\noindent \textbf{Scope}~
%We note that our method is focused strictly on bit-depth quantization and does not consider other strategies often used in bit-depth reduction, such as image halftoning~\cite{sharma:2002:digital}, or color palette reduction, such as in GIF~\cite{GIF}.
%In addition, our method is not attempting inverse-tone mapping~\cite{eilertsen:2017:hdr,Kim2019DeepSJ}, which is a technique for producing high-dynamic-range images from low-dynamic-range input images. While on the surface these appear as the same problem, the objectives of inverse-tone mapping and bit-depth recovery are different. Inverse tone mapping is an image enhancement task that tries to maximize the \emph{subjective} quality (this can include hallucinating details in under-/over-exposed regions) so that the result looks plausible on an HDR display. As previously mentioned, the aim of bit-depth recovery, and our algorithm in particular, is image restoration where the goal is to accurately recover the actual bits lost due to linear quantization.

\begin{figure*}[!t]
%\begin{center}
\centering
\includegraphics[width=0.97\linewidth]{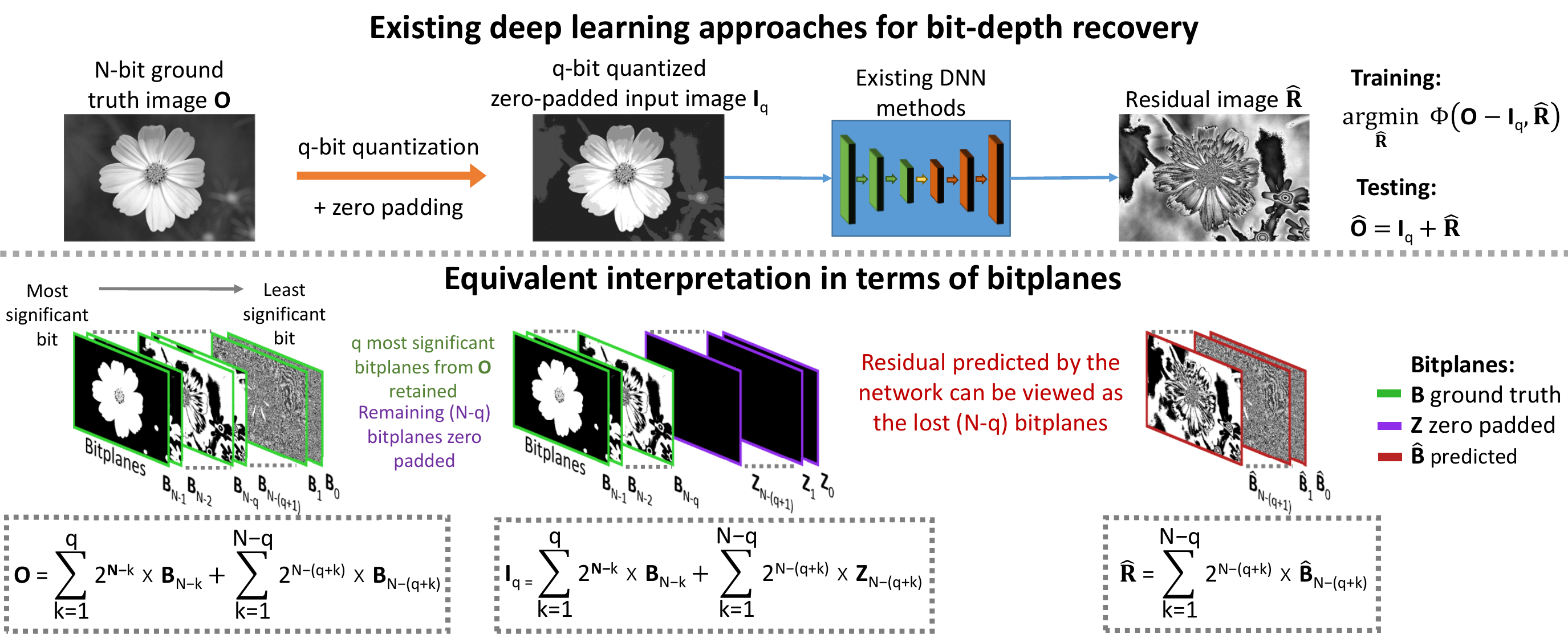}
%\end{center}
\caption{Bit-depth recovery from a bitplane perspective. Existing DNN methods are trained to predict a residual image which when added to the input $q$-bit quantized LBD image produces the desired $N$-bit HBD output image. Interpreted as bitplanes, recovering the residual is equivalent to recovering the $(N$-$q)$ bitplanes lost during quantization. As opposed to existing methods that predict the full residual in a single shot, we propose to recover it one bitplane at a time using a multi-level-supervised training strategy. Specifically, we train $(N$-$q)$ separate networks in a supervised fashion where the input/target pairs for each network are obtained by applying the appropriate level of quantization to the ground truth HBD image.
\label{fig:overview}}
\end{figure*}

%-------------------------------------------------------------------------
\section{Related work}

Early work on bit-depth recovery was based on simple rules to fill the missing bits. Multiplication by an ideal gain is the most straightforward method, wherein the LBD image is simply multiplied by an ideal gain factor. Bit replication~\cite{BR} fills the lost bits with copies of the current LBD bits. Although these methods are fast, they ignore the spatial characteristics of the image and produce visual artifacts. The minimum risk-based classification method~\cite{MRC} constructs prediction error histograms and assigns the value with the minimum associated risk as the HBD pixel intensity at that location. Interpolation-based methods~\cite{CRR,CA} have also been proposed. The contour region reconstruction algorithm~\cite{CRR} linearly interpolates pixel values by analyzing distances from upward contour edges and downward contour edges. A drawback of this method is that it fails in regions with local extrema. The content adaptive technique in~\cite{CA} tackles this issue through the use of a virtual skeleton marking algorithm which converts the problematic 2D extrapolation in local extrema regions into simple 1D interpolation. Optimization-based approaches, such as~\cite{akira,ACDC}, are also popular. With this approach, the bit-depth recovery problem is formulated as a maximum a posteriori (MAP) estimation in~\cite{akira}. The ACDC algorithm~\cite{ACDC} applies graph signal processing to the bit-depth recovery task. The AC component of the desired HBD signal is first calculated using a MAP formulation, and the DC component is then computed using the AC estimate by applying a minimum MSE criterion. The IPAD technique~\cite{IPAD} employs an intensity potential field to model the spatial correlation among the LBD pixels. The HBD image is recovered using a context-adaptive de-quantization procedure that leverages this potential field.

Recently, deep learning-based approaches have gained in popularity~\cite{BE-CNN,BE-CALF,BitNet,BDEN,Hou,GG-DCNN,jing:2018:bde}. Works by Hou and Qiu~\cite{Hou} and GG-DCNN~\cite{GG-DCNN} employ a U-Net style architecture to predict the HBD image given the LBD image. Their methods aim to restore very low bit depth images (e.g., 2 bits). The networks of BE-CNN~\cite{BE-CNN} and Liu et al.~\cite{jing:2018:bde} comprise a chain of deconvolution layers with long and short skip connections. BitNet~\cite{BitNet} uses an encoder-decoder architecture with dilated convolutions and
multi-scale feature integration. BE-CALF~\cite{BE-CALF} employs a chain of convolutional-deconvolutional layers with dense concatetations of all level features. BDEN~\cite{BDEN} uses a two-stream architecture, one for flat and another for non-flat regions, with a local adaptive adjustment preprocessing step for the flat areas. Their approach is tailored towards relatively smaller expansions of the bit depth -- 6 to 8 bit, 8 to 10 bit -- while other methods~\cite{BE-CALF,BE-CNN,BitNet} target larger expansions -- 4 to 16 bit, 6 to 16 bit. Different deep learning methods use different loss functions to train their models; MSE, mean absolute error (MAE), VGG features and so forth are the most common. It can be seen that existing deep learning methods resort to deeper and more complex architectures (U-Net style encoder-decoder, long--short skip connections, multi-stream networks, dense feature concatenations) to improve performance. In the following sections, we demonstrate how we can outperform these existing techniques across a variety of bit-depth ranges using a standard ResNet-style~\cite{ResNet} network architecture. This is made possible by our bitplane-wise training strategy, explained in the following section.

\begin{figure*}[!t]
%\begin{center}
\centering
\includegraphics[width=0.97\linewidth]{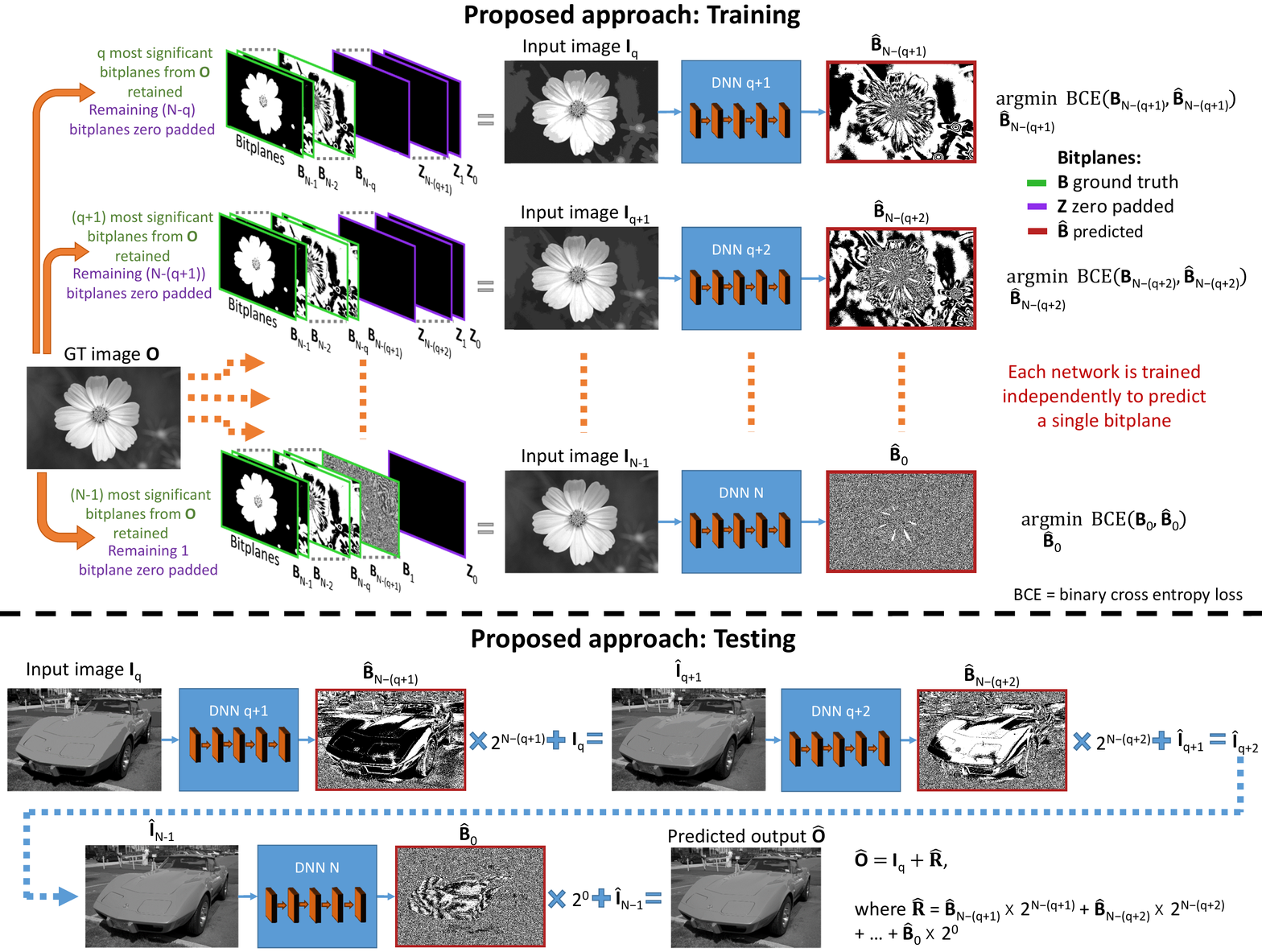}
%\end{center}
\caption{Overview of our proposed approach. We train $(N$-$q)$ separate DNNs to recover the $(N$-$q)$ bitplanes lost during quantization. Here, $N$ is the bit depth of the desired HBD image, and $q$ the bit depth of the input LBD image. The input-target pairs $(\mathbf{I}_{q+k-1},\mathbf{B}_{N-(q+k)})$, $k$=$1$ to $(N$-$q)$, for training each network can be computed directly from the HBD ground truth image $\mathbf{O}$. The target $\mathbf{B}_{N-(q+k)}$ is a binary map, and the problem reduces to one of binary image segmentation, which we optimize using a binary cross entropy loss. At test time, the $q$-bit quantized input LBD image $\mathbf{I}_q$ is passed through the trained networks \textit{sequentially}, with each network increasing the bit depth by one until the image is restored to its desired bit depth of $N$. In particular, each individual network predicts the next bitplane, which is then weighted by the corresponding bit position, and added back to the input image to increment its bit depth by one. This estimated image forms the input to the next network, and so forth.
\label{fig:proposed}}
\end{figure*}

\section{Proposed method}

In this section, we first analyze the bit-depth recovery problem from a bitplane perspective. We then introduce our bitplane-wise training and testing strategy. Our network architecture is discussed in Sec. \ref{sec:arch}.

Let the ground truth HBD image be denoted by $\mathbf{O}$. Assume $\mathbf{O}$ has a bit depth of $N$. To quantize $\mathbf{O}$ to $q$ bits, the following formula can be applied~\cite{Hou,GG-DCNN,li:2005:compressing}:
\begin{eqnarray}
\mathbf{I}_q=\lfloor \frac{\mathbf{O}}{2^{(N-q)}}   \rfloor 2^{(N-q)},
\label{eqn:quant}
\end{eqnarray}
where $\mathbf{I}_q$ represents the $q$-bit quantized and zero padded LBD image, and $\lfloor . \rfloor$ represents the \texttt{floor} operation. Existing deep learning algorithms, such as~\cite{BE-CNN,BitNet,BDEN,Hou,GG-DCNN,jing:2018:bde}, train using ($\mathbf{I}_q$,$\mathbf{O}$) pairs as input and target. We can also express $\mathbf{O}$ as the summation of $\mathbf{I}_q$ and a residual image $\mathbf{R}$:
\begin{eqnarray}
\mathbf{O}=\mathbf{I}_q+\mathbf{R}.
\label{eqn:res_model}
\end{eqnarray}
Liu et al.~\cite{BE-CALF} argue that training using ($\mathbf{I}_q$,$\mathbf{R}$) pairs as input and target yields better results. At test time, the estimate of the residual $\mathbf{\widehat{R}}$ can simply be added to the input $\mathbf{I}_q$ to generate the desired HBD image $\mathbf{\widehat{O}}$. See row one of Fig. \ref{fig:overview}.

Our proposed method is based on the residual model in equation~\eqref{eqn:res_model}.
In row two of Fig. \ref{fig:overview}, we present its equivalent interpretation in terms of bitplanes. The $N$-bit HBD ground truth image $\mathbf{O}$ can be expressed as
\begin{eqnarray}
\mathbf{O} = \sum_{k=1}^q 2^{N-k} \times \mathbf{B}_{N-k} + \sum_{k=1}^{N-q} 2^{N-(q+k)} \times \mathbf{B}_{N-(q+k)},
\end{eqnarray}
where the first term corresponds to the $q$ most significant bits and the second term corresponds to the $(N$-$q)$ least significant bits, and $\mathbf{B}$ denotes a binary map of 0s and 1s. When $\mathbf{O}$ is quantized to $q$ bits using equation~\eqref{eqn:quant}, we get
\begin{eqnarray}
\mathbf{I}_q = \sum_{k=1}^q 2^{N-k} \times \mathbf{B}_{N-k} + \sum_{k=1}^{N-q} 2^{N-(q+k)} \times \mathbf{Z}_{N-(q+k)},
\end{eqnarray}
where all entries of $\mathbf{Z}$ are zero. The residual $\mathbf{\widehat{R}}$ predicted by the network is thus
\begin{eqnarray}
\mathbf{\widehat{R}} =  \sum_{k=1}^{N-q} 2^{N-(q+k)} \times \mathbf{\widehat{B}}_{N-(q+k)},
\end{eqnarray}
where $\mathbf{\widehat{B}}$ is an estimate of the bitplanes. We use grayscale images in Fig. \ref{fig:overview} for ease of visualization of the bitplanes -- the extension to color images is straightforward.
\begin{figure*}
%\begin{center}
\centering
\includegraphics[width=0.7\linewidth]{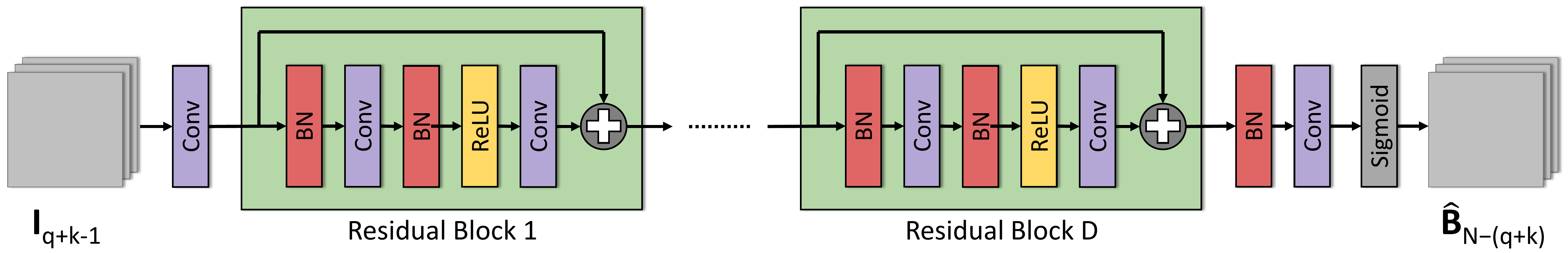}
%\end{center}
\caption{A diagram of our network architecture. We train $(N$-$q)$ networks, one for each bitplane lost during quantization. Compared to existing DNN architectures for bit-depth recovery that use encoder-decoder modules, dense feature concatenation layers, multiple streams, and more, our architecture is a standard ResNet-style~\cite{ResNet} structure with a series of residual blocks. Even with just four residual blocks (i.e., $D$=$4$), we can achieve state-of-the-art results based on our bitplane-wise training framework.
\label{fig:network}}
\end{figure*}

Next, we describe in detail our training strategy. Given a $q$-bit quantized image $\mathbf{I}_q$, the objective is to recover the $(N$-$q)$ bitplanes $\mathbf{\widehat{B}}$ lost during quantization, and restore $\mathbf{I}_q$ to its original bit depth of $N$. Towards this goal, we train $(N$-$q)$ networks independently as shown in Fig. \ref{fig:proposed}, where each network is trained to predict one single bitplane. Observe that $\mathbf{I}_q$ retains $q$ most significant bits from $\mathbf{O}$. Therefore, the first network labelled DNN $(q$+$1)$ is trained to predict the bitplane $N$-$(q$+$1)$. The input and target pairs for this network are the $q$-bit quantized image $\mathbf{I}_q$ and the ground truth binary map $\mathbf{B}_{N-(q+1)}$, respectively, both of which can be obtained from the ground truth HBD image $\mathbf{O}$. We use the binary cross entropy (BCE) loss, which is typically applied to binary image segmentation problems, to train our network. The loss is computed between the ground truth binary map $\mathbf{B}_{N-(q+1)}$ and the network's prediction $\mathbf{\widehat{B}}_{N-(q+1)}$.

In a similar manner, to predict the next least significant bit, we train DNN $(q+2)$ using the input-target pair $(\mathbf{I}_{q+1},\mathbf{B}_{N-(q+2)})$, where $\mathbf{I}_{q+1}$ is obtained by quantizing $\mathbf{O}$ to $(q+1)$ bits using equation~\eqref{eqn:quant}. This process is repeated for all $(N$-$q)$ bits -- the last least significant bitplane is trained using DNN $(N)$ with an $(N$-$1)$-bit quantized image $\mathbf{I}_{N-1}$ as input, and the ground truth binary map $\mathbf{B}_0$ as target. All our networks are trained using the same BCE loss as shown in Fig. \ref{fig:proposed}.
\begin{table*}[]
\caption{Results on Sintel dataset~\cite{Sintel}. The best results are reported in bold and red. The second, third, and fourth best-performing methods are shown in green, blue, and yellow,
respectively. Numbers copied directly from the original papers are marked with a $\dagger$ symbol.
\label{tab:Sintel_16_bit_validation}}
\centering
%\begin{center}
\scriptsize
\setlength{\tabcolsep}{4pt}
\begin{tabular}{ccccccccccccccc}
\toprule
\multicolumn{2}{c}{} & ZP & MIG & BR & MRC & CRR & CA & ACDC & IPAD & BE-CNN & BE-CALF & BitNet & Ours D4 & Ours D16 \\
\multicolumn{2}{c}{} & & & \cite{BR} & \cite{MRC} & \cite{CRR} & \cite{CA} & \cite{ACDC} & \cite{IPAD} & \cite{BE-CNN} & \cite{BE-CALF} & \cite{BitNet} & & \\
\toprule
\multirow{2}{*}{4-16 bit} & PSNR & 28.7692 & 31.9004 & 32.4604 & 33.7792 & 33.7982 & 35.5001 & 34.6394 & 35.7647 & 35.7137 & \cellcolor{blue!25} 39.9829 & \cellcolor{yellow!75} 39.4893 & \cellcolor{green!25} 40.9274 & \cellcolor{red!25} \textbf{41.5070} \\

 	 	 	 	 	 	  & SSIM & 0.8843 & 0.8826 & 0.8947 & 0.9126 & 0.9348 & 0.9436 & 0.9077 & 0.9451 & 0.9578 & \cellcolor{blue!25} 0.9752 & \cellcolor{yellow!75} 0.9719 & \cellcolor{green!25} 0.9786 & \cellcolor{red!25} \textbf{0.9810} \\
\midrule
\multirow{2}{*}{4-12 bit} & PSNR & 28.7916 & 31.9140 & 32.4655 & 33.7915 & 33.8342 & 35.5171 & 34.6384 & 35.7753 & 35.7136 & \cellcolor{blue!25} 39.9840 & \cellcolor{yellow!75} 39.4931 & \cellcolor{green!25} 40.9286 & \cellcolor{red!25} \textbf{41.5080} \\

 	 	 	 	 	 	  & SSIM & 0.8844 & 0.8827 & 0.8948 & 0.9126 & 0.9352 & 0.9438 & 0.9077 & 0.9452 & 0.9578 & \cellcolor{blue!25} 0.9752 & \cellcolor{yellow!75} 0.9719 & \cellcolor{green!25} 0.9786 & \cellcolor{red!25} \textbf{0.9810} \\
\midrule
\multirow{2}{*}{4-8 bit} & PSNR & 29.1599 & 32.1404 & 32.6690 & 33.9525 & 34.3592 & 35.7051 & 34.5944 & 35.8610 & 35.6839 & \cellcolor{blue!25} 39.9072 & \cellcolor{yellow!75} 39.3369 & \cellcolor{green!25} 40.6143 & \cellcolor{red!25} \textbf{41.1909} \\

 	 	 	 	 	 	  & SSIM & 0.8864 & 0.8847 & 0.8989 & 0.9141 & 0.9389 & 0.9444 & 0.9074 & 0.9457 & 0.9566 & \cellcolor{blue!25} 0.9737 & \cellcolor{yellow!75} 0.9701 & \cellcolor{green!25} 0.9773 & \cellcolor{red!25} \textbf{0.9794} \\
\midrule
\multirow{2}{*}{6-16 bit} & PSNR & 40.8072 & 44.2780 & 44.4131 & 46.8504 & 46.0178 & 46.9613 & 46.6553 & 47.6154 & \cellcolor{yellow!75} 49.7405$^{\dagger}$ & \cellcolor{blue!25} 51.1430$^{\dagger}$ & 49.6795 & \cellcolor{green!25} 52.7599 & \cellcolor{red!25} \textbf{53.4825} \\

 	 	 	 	 	 	  & SSIM & 0.9857 & 0.9858 & 0.9862 & 0.9903 & 0.9864 & 0.9896 & 0.9858 & 0.9902 & 0.9926$^{\dagger}$ & \cellcolor{yellow!75} 0.9940$^{\dagger}$ & \cellcolor{blue!25} 0.9954 & \cellcolor{green!25} 0.9976 & \cellcolor{red!25} \textbf{0.9979} \\
\midrule
\multirow{2}{*}{6-12 bit} & PSNR & 40.9029 & 44.3357 & 44.4725 & 46.8886 & 46.1370 & 47.0376 & 46.6522 & 47.6593 & \cellcolor{yellow!75} 49.7421$^{\dagger}$ & \cellcolor{blue!25} 51.1454$^{\dagger}$ & 49.7192 & \cellcolor{green!25} 52.7491 & \cellcolor{red!25} \textbf{53.4731} \\

 	 	 	 	 	 	  & SSIM & 0.9858 & 0.9858 & 0.9864 & 0.9903 & 0.9867 & 0.9898 & 0.9858 & 0.9903 & 0.9926$^{\dagger}$ & \cellcolor{yellow!75} 0.9940$^{\dagger}$ & \cellcolor{blue!25} 0.9954 & \cellcolor{green!25} 0.9976 & \cellcolor{red!25} \textbf{0.9980} \\
\midrule
\multirow{2}{*}{8-16 bit} & PSNR & 52.8604 & 56.3970 & 56.4317 & \cellcolor{yellow!75} 59.3085 & 57.4125 & 57.8523 & 58.6982 & 58.6227 & 54.7790 & \cellcolor{blue!25} 59.5117 & 57.5487 & \cellcolor{green!25} 63.0731 & \cellcolor{red!25} \textbf{63.5146} \\

 	 	 	 	 	 	  & SSIM & 0.9990 & 0.9990 & 0.9990 & \cellcolor{yellow!75} 0.9993 & 0.9981 & 0.9988 & 0.9989 & 0.9989 & 0.9989 & \cellcolor{blue!25} 0.9993 & 0.9989 & \cellcolor{green!25} 0.9997 & \cellcolor{red!25} \textbf{0.9998} \\
\bottomrule
\end{tabular}
%\end{center}
\end{table*}

At test time, the trained networks are applied sequentially as shown in Fig. \ref{fig:proposed}. The $q$-bit quantized test image $\mathbf{I}_q$ is first fed to DNN $(q+1)$. The network's prediction $\mathbf{\widehat{B}}_{N-(q+1)}$ is multiplied by the appropriate weighting factor $2^{N-(q+1)}$ and this result is added to the input image $\mathbf{I}_q$ to obtain our estimate $\mathbf{\widehat{I}}_{q+1}$. Note that at this stage, $\mathbf{\widehat{I}}_{q+1}$, we have restored the $q$-bit quantized input $\mathbf{I}_q$ to a bit depth of $(q+1)$. Next, we input this estimate $\mathbf{\widehat{I}}_{q+1}$ to DNN $(q+2)$ to obtain its prediction $\mathbf{\widehat{B}}_{N-(q+2)}$, which is multiplied by $2^{N-(q+2)}$ and added to  $\mathbf{\widehat{I}}_{q+1}$ to obtain our $(q+2)$-bit-depth estimate, $\mathbf{\widehat{I}}_{q+2}$. Repeating this process till the last bit produces our estimate of the HBD image $\mathbf{\widehat{O}}$.

\subsection{Network architecture}
\label{sec:arch}
Our network architecture is shown in Fig. \ref{fig:network}. The input is a 3-channel RGB image $\mathbf{I}_{q+k-1}$. The network predicts a 3-channel output  $\mathbf{\widehat{B}}_{N-(q+k)}$ having the same size as the input. Here, $k$ is the index that runs over the bitplanes -- namely, $k$=$1$ to $(N$-$q)$. We use the same architecture for all $(N$-$q)$ bitplanes. As previously mentioned, each of these $(N$-$q)$ networks is trained independently. Our architecture consists of an initial convolution layer (Conv), a series of $D$ residual blocks, followed by a batch normalization (BN) layer and a final convolution layer. The last layer is a sigmoid activation restricting all pixel values to lie in the range [0,1]. Our residual blocks are styled after~\cite{BDEN} and contain two convolution layers, two batch normalization layers, and a rectified linear unit (ReLU). There is an additive skip connection between the input and output of each residual block. All convolution filters are of size $3 \times 3$ and padded such that the input and output are of the same size. We use a fixed number of 64 filters for all convolution layers. The depth of our network is determined primarily by the number of residual units $D$. As we shall demonstrate in Sec. \ref{sec:results}, we can already outperform all competing approaches with $D$=$4$. We also experiment with $D$=$16$ and show that this gives up a further boost in performance.

We note that our primary focus is \emph{not} the choice of network architecture, but rather to evaluate the effectiveness of our bitplane-wise training strategy for bit-depth recovery. Other recent works have used more complex architectures, such as dense feature concatenations, multiple streams, and encoder-decoder structures. While such configurations are indeed possible, our aim is to demonstrate that a standard ResNet-style~\cite{ResNet} architecture can produce state-of-the-art results with bitplane-wise training. We would also like to add that while our architecture resembles a single stream of BDEN~\cite{BDEN}, their overall network structure is more complex with two streams and a local adaptive adjustment preprocessing step applied to one stream.

Our method requires training separate models for each bitplane. While this may be seen as cumbersome, competing methods, such as~\cite{BDEN,Hou}, also train separate models depending on the bit depth of the quantized input image.

\section{Results}
\label{sec:results}

In this section, we first provide details regarding how we train our networks. Next, we evaluate the performance of our algorithm against competing approaches both quantitatively and qualitatively using five publicly available image datasets.

\subsection{Training}
We use 2000 color images, 1000 each from the Sintel dataset~\cite{Sintel} and the MIT-Adobe FiveK dataset~\cite{Adobe}, for training. Sintel is a short animation film, and the dataset contains more than 20,000 16-bit lossless images at a resolution of $436 \times 1024$ pixels. Following~\cite{BE-CALF,BE-CNN}, we select 1000 images at random from this dataset. The MIT-Adobe FiveK dataset is a natural image dataset consisting of 5000 images in raw format captured using different SLR cameras covering a broad range of scenes, subjects, and lighting conditions. Each raw image has been retouched in Adobe Lightroom by five professionals, and these five variations, saved as 16-bit lossless TIFF files, are available as part of the dataset. We use the first 1000 images (i.e., filenames a0001 to a1000) enhanced by a random expert for training. To ensure that images across both datasets are roughly the same resolution, all images from the MIT-Adobe dataset are downsampled by a factor of 4. While some papers~\cite{BDEN,Hou,GG-DCNN,BitNet} use natural images for training, others~\cite{BE-CALF,BE-CNN} argue that animated images represent harder examples, and are thus more appropriate. As compared to training on only one category of images, we found that a balanced mix of animated and natural images gives the best generalization across both animated and natural image datasets.
%In our experiments, we found that a balanced mix of animated and natural images gives the best generalization.
\begin{table}[!b]
\caption{Results on MIT-Adobe FiveK dataset~\cite{Adobe}.
\label{tab:Adobe_1K_test}}
%\begin{center}
\centering
\scriptsize
\setlength{\tabcolsep}{2pt}
\begin{tabular}{ccccccccc}
\toprule
\multicolumn{2}{c}{} & ZP & MIG & BR & IPAD & BitNet & Ours D4 & Ours D16 \\
\multicolumn{2}{c}{} & & & \cite{BR} & \cite{IPAD} & \cite{BitNet} & & \\
\toprule
\multirow{2}{*}{3-16 bit} & PSNR & 22.8986 & 25.2658 & 26.4256 & \cellcolor{yellow!75} 29.8615 & \cellcolor{blue!25} 33.4621 & \cellcolor{green!25} 33.7891 & \cellcolor{red!25} \textbf{34.1088} \\

 	 	 	 	 	 	  & SSIM & 0.7381 & 0.7334 & 0.7817 & \cellcolor{yellow!75} 0.8624 & \cellcolor{blue!25} 0.9128 & \cellcolor{green!25} 0.9252 & \cellcolor{red!25} \textbf{0.9279} \\
\midrule
\multirow{2}{*}{4-16 bit} & PSNR & 28.8587 & 31.7121 & 32.2720 & \cellcolor{yellow!75} 35.7357 & \cellcolor{blue!25} 39.2103 & \cellcolor{green!25} 39.6484 & \cellcolor{red!25} \textbf{39.9478} \\

 	 	 	 	 	 	  & SSIM & 0.8769 & 0.8745 & 0.8876 & \cellcolor{yellow!75} 0.9378 & \cellcolor{blue!25} 0.9632 & \cellcolor{green!25} 0.9683 & \cellcolor{red!25} \textbf{0.9693} \\
\midrule
\multirow{2}{*}{5-16 bit} & PSNR & 34.8596 & 37.9655 & 38.2413 & \cellcolor{yellow!75} 41.1847 & \cellcolor{blue!25} 44.0214 & \cellcolor{green!25} 44.8035 & \cellcolor{red!25} \textbf{44.9396} \\

 	 	 	 	 	 	  & SSIM & 0.9556 & 0.9554 & 0.9580 & \cellcolor{yellow!75} 0.9743 & \cellcolor{blue!25} 0.9853 & \cellcolor{green!25} 0.9871 & \cellcolor{red!25} \textbf{0.9876} \\
\midrule
\multirow{2}{*}{6-16 bit} & PSNR & 40.8795 & 44.1029 & 44.2380 & \cellcolor{yellow!75} 46.4284 & \cellcolor{blue!25} 48.4657 & \cellcolor{green!25} 49.5657 & \cellcolor{red!25} \textbf{49.7193} \\

 	 	 	 	 	 	  & SSIM & 0.9871 & 0.9872 & 0.9876 & \cellcolor{yellow!75} 0.9903 & \cellcolor{blue!25} 0.9943 & \cellcolor{green!25} 0.9951 & \cellcolor{red!25} \textbf{0.9953} \\
\bottomrule
\end{tabular}
%\end{center}
\end{table}

\begin{table*}[]
\caption{Results on TESTIMAGES 1200 dataset~\cite{USTHK}.
\label{tab:UST_HK_1200}}
%\begin{center}
\centering
\scriptsize
\setlength{\tabcolsep}{4pt}
\begin{tabular}{ccccccccccccccc}
\toprule
\multicolumn{2}{c}{} & ZP & MIG & BR & MRC & CRR & CA & ACDC & IPAD & BE-CNN & BE-CALF & BitNet & Ours D4 & Ours D16 \\
\multicolumn{2}{c}{} & & & \cite{BR} & \cite{MRC} & \cite{CRR} & \cite{CA} & \cite{ACDC} & \cite{IPAD} & \cite{BE-CNN} & \cite{BE-CALF} & \cite{BitNet} & \\
\toprule

\multirow{2}{*}{4-16 bit} & PSNR & 28.8322 & 31.5393 & 32.0988 & 34.2227 & 33.5094 & 35.1968 & 34.7447 & 36.1890 & 32.3203 & \cellcolor{yellow!75} 38.5099 & \cellcolor{blue!25} 38.8073 & \cellcolor{green!25} 39.6503 & \cellcolor{red!25} \textbf{40.4099} \\ 

 	 	 	 	 	 	  & SSIM & 0.8739 & 0.8711 & 0.8845 & 0.9169 & 0.9243 & 0.9343 & 0.8994 & 0.9443 & 0.9418 & \cellcolor{blue!25} 0.9649 & \cellcolor{yellow!75} 0.9589 & \cellcolor{green!25} 0.9700 & \cellcolor{red!25} \textbf{0.9735} \\ 
\midrule
\multirow{2}{*}{4-12 bit} & PSNR & 28.8543 & 31.5545 & 32.0993 & 34.2385 & 33.5428 & 35.2121 & 34.7422 & 36.2000 & 32.3191 & \cellcolor{yellow!75} 38.5095 & \cellcolor{blue!25} 38.8158 & \cellcolor{green!25} 39.6619 & \cellcolor{red!25} \textbf{40.4216} \\ 

 	 	 	 	 	 	  & SSIM & 0.8741 & 0.8712 & 0.8847 & 0.9170 & 0.9247 & 0.9344 & 0.8993 & 0.9444 & 0.9417 & \cellcolor{blue!25} 0.9648 & \cellcolor{yellow!75} 0.9589 & \cellcolor{green!25} 0.9700 & \cellcolor{red!25} \textbf{0.9735} \\ 
\midrule
\multirow{2}{*}{4-8 bit} & PSNR & 29.2192 & 31.8072 & 32.2102 & 34.4698 & 34.0535 & 35.3879 & 34.6727 & 36.2924 & 32.2774 & \cellcolor{yellow!75} 38.4572 & \cellcolor{blue!25} 38.7515 & \cellcolor{green!25} 39.6822 & \cellcolor{red!25} \textbf{40.3906} \\ 

 	 	 	 	 	 	  & SSIM & 0.8764 & 0.8736 & 0.8896 & 0.9175 & 0.9295 & 0.9354 & 0.8986 & 0.9450 & 0.9403 & \cellcolor{blue!25} 0.9632 & \cellcolor{yellow!75} 0.9571 & \cellcolor{green!25} 0.9691 & \cellcolor{red!25} \textbf{0.9725} \\ 
\midrule

\multirow{2}{*}{6-16 bit} & PSNR & 40.8621 & 43.8101 & 43.9437 & 47.0584 & 45.3877 & 45.2110 & 46.7708 & 47.1574 & 46.9513$^{\dagger}$ & \cellcolor{blue!25} 49.8488$^{\dagger}$ & \cellcolor{yellow!75} 49.4834 & \cellcolor{green!25} 51.5413 & \cellcolor{red!25} \textbf{52.1204} \\

 	 	 	 	 	 	  & SSIM & 0.9855 & 0.9856 & 0.9861 & 0.9912 & 0.9852 & 0.9881 & 0.9871 & 0.9899 & 0.9924$^{\dagger}$ & \cellcolor{blue!25} 0.9945$^{\dagger}$ & \cellcolor{yellow!75} 0.9944 & \cellcolor{green!25} 0.9964 & \cellcolor{red!25} \textbf{0.9967} \\
\midrule
\multirow{2}{*}{6-12 bit} & PSNR & 40.9570 & 43.8747 & 43.9823 & 47.0986 & 45.5076 & 45.2845 & 46.7661 & 47.2052 & 46.9528$^{\dagger}$ & \cellcolor{blue!25} 49.8521$^{\dagger}$ & \cellcolor{yellow!75} 49.5259 & \cellcolor{green!25} 51.5490 & \cellcolor{red!25} \textbf{52.1220} \\

 	 	 	 	 	 	  & SSIM & 0.9856 & 0.9857 & 0.9863 & 0.9912 & 0.9856 & 0.9883 & 0.9871 & 0.9901 & 0.9924$^{\dagger}$ & \cellcolor{blue!25} 0.9945$^{\dagger}$ & \cellcolor{yellow!75} 0.9944 & \cellcolor{green!25} 0.9964 & \cellcolor{red!25} \textbf{0.9967} \\

\midrule
\multirow{2}{*}{8-16 bit} & PSNR & 52.9243 & 55.9065 & 55.9430 & \cellcolor{blue!25} 59.0353 & 56.8642 & 55.4075 & \cellcolor{yellow!75} 58.8097 & 57.8428 & 53.1379 & 58.1167 & 53.6031 & \cellcolor{green!25} 61.3626 & \cellcolor{red!25} \textbf{61.6839} \\ 

 	 	 	 	 	 	  & SSIM & 0.9990 & 0.9990 & 0.9990 & \cellcolor{blue!25} 0.9993 & 0.9982 & 0.9986 & 0.9991 & 0.9988 & 0.9986 & \cellcolor{yellow!75} 0.9992 & 0.9970 & \cellcolor{green!25} 0.9996 & \cellcolor{red!25} \textbf{0.9996}  \\ 

\bottomrule
\end{tabular}
%\end{center}
\end{table*}

\begin{table*}[]
\caption{Results on Kodak dataset~\cite{Kodak}. NR denotes that a score was `not reported' in the original paper.
\label{tab:Kodak_data}}
%\begin{center}
\centering
\scriptsize
\setlength{\tabcolsep}{4pt}
\begin{tabular}{ccccccccccccccc}
\toprule
\multicolumn{2}{c}{} & ZP & MIG & BR & MRC & CRR & CA & ACDC & IPAD & BE-CNN & BE-CALF & BitNet & Ours D4 & Ours D16 \\
\multicolumn{2}{c}{} & & & \cite{BR} & \cite{MRC} & \cite{CRR} & \cite{CA} & \cite{ACDC} & \cite{IPAD} &  \cite{BE-CNN} & \cite{BE-CALF} & \cite{BitNet} & & \\
\toprule
\multirow{2}{*}{3-8 bit} & PSNR & 22.7767 & 25.8250 & 27.0293 & 28.3804 & 28.2246 & 29.1447 & 28.6566 & \cellcolor{yellow!75} 29.2012 & NR & NR & \cellcolor{blue!25} 32.6832 & \cellcolor{green!25} 33.5089 & \cellcolor{red!25} \textbf{33.6679} \\

 	 	 	 	 	 	  & SSIM & 0.7671 & 0.7604 & 0.8036 & 0.8246 & 0.8304 & 0.8413 & 0.8200 & \cellcolor{yellow!75} 0.8515 & NR & NR & \cellcolor{blue!25} 0.9172 & \cellcolor{green!25} 0.9319 & \cellcolor{red!25} \textbf{0.9337} \\
\midrule
\multirow{2}{*}{4-8 bit} & PSNR & 29.0657 & 32.6293 & 33.3027 & 35.2607 & 34.1294 & 34.7382 & 34.6817 & 34.9081 & 35.0585 & \cellcolor{blue!25} 38.9271 & \cellcolor{yellow!75} 38.4822 & \cellcolor{green!25} 39.4171 & \cellcolor{red!25} \textbf{39.5185} \\

 	 	 	 	 	 	  & SSIM & 0.8998 & 0.8969 & 0.9108 & 0.9270 & 0.9293 & 0.9317 & 0.9152 & 0.9345 & 0.9575 & \cellcolor{blue!25} 0.9681 & \cellcolor{yellow!75} 0.9659 & \cellcolor{green!25} 0.9709 & \cellcolor{red!25} \textbf{0.9723} \\
\bottomrule
\end{tabular}
%\end{center}
\end{table*}

\begin{table*}[]
\caption{Results on ESPL v2 dataset~\cite{ESPL}.
\label{tab:ESPL_v2_8_bit}}
%\begin{center}
\centering
\scriptsize
\setlength{\tabcolsep}{4pt}
\begin{tabular}{ccccccccccccccc}
\toprule
\multicolumn{2}{c}{} & ZP & MIG & BR & MRC & CRR & CA & ACDC & IPAD & BE-CNN & BE-CALF & BitNet & Ours D4 & Ours D16 \\
\multicolumn{2}{c}{} & & & \cite{BR} & \cite{MRC} & \cite{CRR} & \cite{CA} & \cite{ACDC} & \cite{IPAD} & \cite{BE-CNN} & \cite{BE-CALF} & \cite{BitNet} & & \\
\toprule
\multirow{2}{*}{3-8 bit} & PSNR & 23.2092 & 25.4391 & 26.6110 & 27.3040 & 26.9249 & 29.4643 & 28.6803 & \cellcolor{yellow!75} 29.8653 & NR & NR & \cellcolor{blue!25} 32.5878 & \cellcolor{green!25} 33.1244 & \cellcolor{red!25} \textbf{33.4685} \\

 	 	 	 	 	 	  & SSIM & 0.6616 & 0.6571 & 0.7242 & 0.7381 & 0.7990 & 0.8245 & 0.7764 & \cellcolor{yellow!75} 0.8379 & NR & NR & \cellcolor{blue!25} 0.8717 & \cellcolor{green!25} 0.8981 & \cellcolor{red!25} \textbf{0.9001} \\
\midrule
\multirow{2}{*}{4-8 bit} & PSNR & 29.2893 & 31.8492 & 32.4288 & 34.2636 & 34.2817 &  35.7807 & 34.6381 & 35.7558 & 32.6545 & \cellcolor{blue!25} 38.4307 & \cellcolor{yellow!75} 38.2329 & \cellcolor{green!25} 39.3854 & \cellcolor{red!25} \textbf{39.5312} \\

 	 	 	 	 	 	  & SSIM & 0.8261 & 0.8240 & 0.8453 & 0.8763 & 0.9046 & 0.9184 & 0.8818 & 0.9207 &  0.9193 & \cellcolor{blue!25} 0.9479 & \cellcolor{yellow!75} 0.9399 & \cellcolor{red!25} \textbf{0.9532} & \cellcolor{green!25} 0.9528 \\
\bottomrule
\end{tabular}
%\end{center}
\end{table*}

For training, we crop non-overlapping patches of size $48 \times 48$ pixels from all 2000 images. Data augmentation is performed by randomly flipping the patches, or rotating them by $90^\circ$. We use the Adam optimizer~\cite{Adam} with parameters $\beta_1=0.9$ and $\beta_2=0.999$, a learning rate of $10^{-3}$, and a batch size of 128. The learning rate is divided by 5 after 16 epochs. All our models are trained for 30 epochs using these same settings. Our method is implemented using TensorFlow~\cite{tensorflow}. Our code and corresponding pre-trained models are publicly available at \url{https://github.com/abhijithpunnappurath/a-little-bit-more}.
%All our trained models and training code will be made publicly available.

\subsection{Evaluation}
We evaluate the performance of our method on three natural image datasets -- MIT-Adobe FiveK~\cite{Adobe}, TESTIMAGES~\cite{USTHK}, and Kodak~\cite{Kodak} -- and two animated image datasets -- Sintel~\cite{Sintel} and ESPL v2~\cite{ESPL}. For quantitative evaluation, PSNR (dB) and structural similarity index (SSIM)~\cite{SSIM} are chosen as metrics. We compare our proposed method against eight classical (i.e., non-deep learning) algorithms -- (1) zero padding (ZP), (2) multiplication by ideal gain (MIG), (3) bit replication (BR)~\cite{BR}, (4) minimum risk-based classification (MRC)~\cite{MRC}, (5) contour region reconstruction (CRR)~\cite{CRR}, (6) content-adaptive image bit-depth enhancement (CA)~\cite{CA},
(7) maximum a posteriori estimation of AC signal (ACDC)~\cite{ACDC}, and (8) intensity potential for image de-quantization (IPAD)~\cite{IPAD}.  We also compare against three recent DNN methods -- (1) bit-depth enhancement with CNN (BE-CNN)~\cite{BE-CNN}, (2) concatenating all level features (BE-CALF)~\cite{BE-CALF}, and (3) learning-based bit-depth expansion (BitNet)~\cite{BitNet}. The codes of the eight classical algorithms are implemented in Matlab\footnote{\url{https://sites.google.com/site/jingliu198810/publication}} by the authors of \cite{IPAD}. For the DNN methods~\cite{BE-CNN,BE-CALF}, pre-trained models have been released by the authors for only two quantization levels -- 4 to 16, and 8 to 16 bit recovery. Training code is not available. Hence, we report results for 4-bit and 8-bit quantized input images using their pre-trained models. For other quantization levels, we copy the results directly from the original papers, where available. For comparison against BitNet~\cite{BitNet}, we report scores using the code released by the authors.

We first report results on the Sintel~\cite{Sintel} dataset. Following~\cite{BE-CALF,BE-CNN}, we select 50 random images (different from the training set) from this dataset for evaluation. Table \ref{tab:Sintel_16_bit_validation} shows the scores of our proposed algorithm as well as various competing methods. Two variants of our proposed method corresponding to two different network depths are tested -- D4 and D16, corresponding to $D$=$4$ and $D$=$16$ residual units, respectively, as explained in Sec. \ref{sec:arch}. The best results are reported in bold and red. The second, third, and fourth best-performing methods are shown in green, blue, and yellow, respectively. Both~\cite{BE-CNN} and~\cite{BE-CALF} are trained exclusively on animated images from Sintel while we use a mix of natural and animated images for training. Yet, as can be seen from the results, both our models outperform all competing methods, including  BE-CALF~\cite{BE-CALF} and BE-CNN~\cite{BE-CNN}, by a sound margin across all quantization levels.

BitNet~\cite{BitNet} has used the last 1000 images (i.e., filenames a4001 to a5000) of the MIT-Adobe dataset~\cite{Adobe} enhanced by expert E for testing. The scores on these images are reported in Table \ref{tab:Adobe_1K_test}. Note that BitNet is trained exclusively on the first 4000 images enhanced by expert E, while we use only 1000 images by a random expert. Once again, it can be observed from the results that both our models outperform all comparison techniques, including BitNet.
\begin{figure}[!t]
\begin{center}
\setlength{\tabcolsep}{2pt}
\begin{tabular}{cc}
\includegraphics[height=0.275\linewidth]{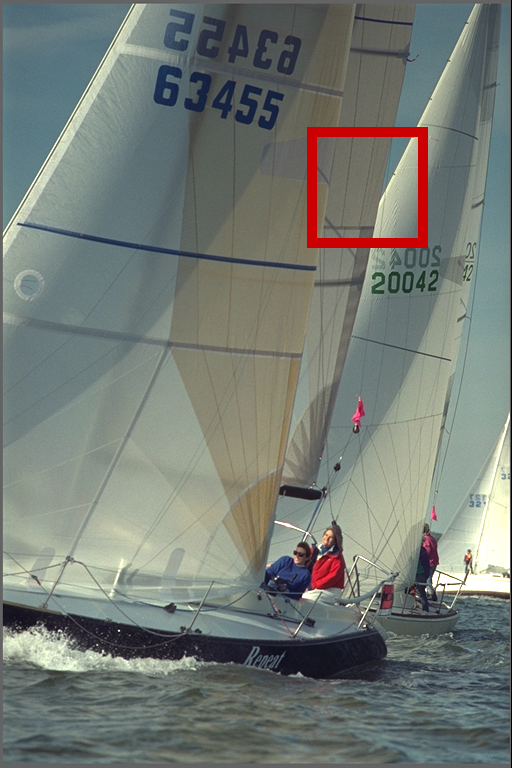}&
\includegraphics[height=0.275\linewidth]{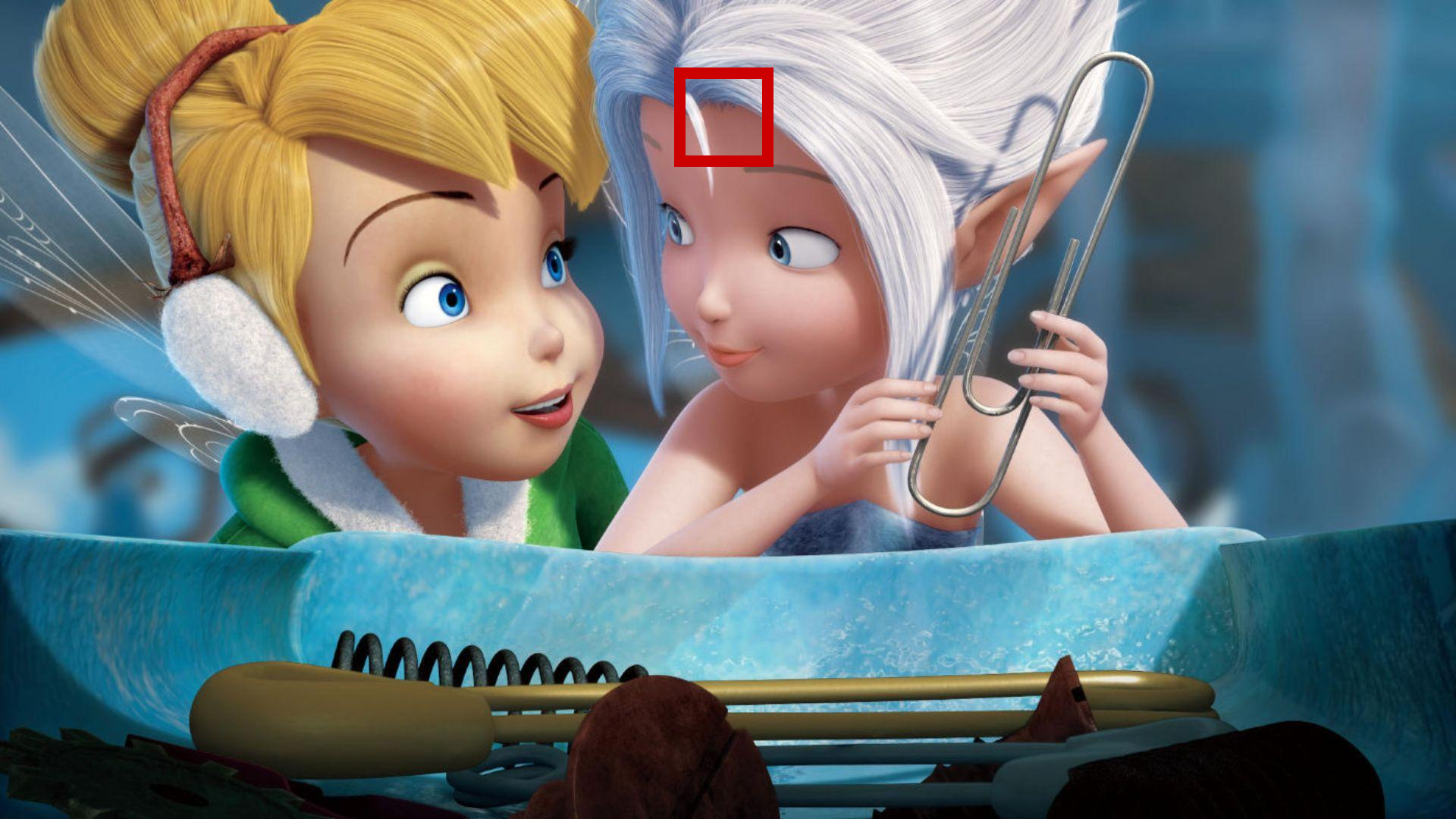}
\end{tabular}
\\ Ground truth (GT)
\begin{tabular}{ccccc}
\includegraphics[width=0.18\linewidth]{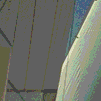}&
\includegraphics[width=0.18\linewidth]{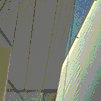}&
\includegraphics[width=0.18\linewidth]{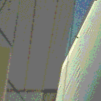}&
\includegraphics[width=0.18\linewidth]{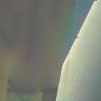}&
\includegraphics[width=0.18\linewidth]{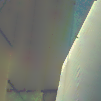}\\
ZP & BR~\cite{BR} & MRC~\cite{MRC} & CRR~\cite{CRR} & CA~\cite{CA} \\

\includegraphics[width=0.18\linewidth]{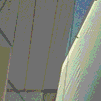}&
\includegraphics[width=0.18\linewidth]{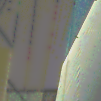}&
\includegraphics[width=0.18\linewidth]{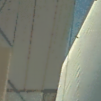}&
\includegraphics[width=0.18\linewidth]{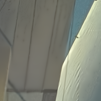}&
\includegraphics[width=0.18\linewidth]{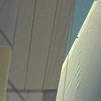}\\
ACDC~\cite{ACDC} & IPAD~\cite{IPAD} & BitNet~\cite{BitNet} & Ours D16 & GT\\

\includegraphics[width=0.18\linewidth]{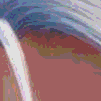}&
\includegraphics[width=0.18\linewidth]{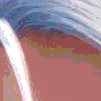}&
\includegraphics[width=0.18\linewidth]{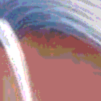}&
\includegraphics[width=0.18\linewidth]{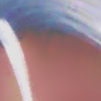}&
\includegraphics[width=0.18\linewidth]{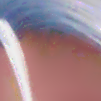}\\
ZP & BR~\cite{BR} & MRC~\cite{MRC} & CRR~\cite{CRR} & CA~\cite{CA} \\

\includegraphics[width=0.18\linewidth]{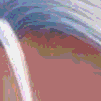}&
\includegraphics[width=0.18\linewidth]{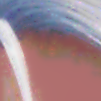}&
\includegraphics[width=0.18\linewidth]{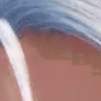}&
\includegraphics[width=0.18\linewidth]{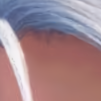}&
\includegraphics[width=0.18\linewidth]{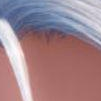}\\
ACDC~\cite{ACDC} & IPAD~\cite{IPAD} & BitNet~\cite{BitNet} & Ours D16 & GT\\
%ZP & MIG & BR~\cite{BR} & MRC~\cite{MRC} & CRR~\cite{CRR} \\
%CA~\cite{CA} & ACDC~\cite{ACDC} & IPAD~\cite{IPAD} & Ours D16 & GT\\
\end{tabular}
\end{center}
\caption{Qualitative comparisons on the natural image Kodak dataset~\cite{Kodak} and the animated image ESPL v2 dataset~\cite{ESPL} for 3 to 8 bit recovery.
\label{fig:qualitative}}
\end{figure}

The remaining three datasets, TESTIMAGES~\cite{USTHK}, Kodak~\cite{Kodak}, and ESPL v2~\cite{ESPL}, are not part of our training set. We evaluate our method's generalizability by testing on these images. The TESTIMAGES dataset~\cite{USTHK} contains 40 $1200 \times 1200$ 16-bit natural images. The scores are reported in Table \ref{tab:UST_HK_1200}. The Kodak dataset~\cite{Kodak} is a natural image dataset consisting of 24 $768 \times 512$ images with a bit depth of 8. The results on this dataset are reported in Table \ref{tab:Kodak_data}. ESPL v2~\cite{ESPL} contains 25 $1920 \times 1080$ animated images also with a bit depth of 8. Following~\cite{BitNet}, we used only the pristine images without any distortion, and the scores are presented in Table \ref{tab:ESPL_v2_8_bit}.
%Lastly, we compare against Hou et al.~\cite{Hou} and GG-DCNN~\cite{GG-DCNN}, who have trained and tested their models on the Microsoft COCO~\cite{MSCOCO} dataset. Following GG-DCNN~\cite{GG-DCNN}, we use 2000 random images from the testing fold of this dataset. The authors of ~\cite{GG-DCNN} have re-implemented the method of~\cite{Hou}, and in Table \ref{tab:MSCOCO_2K}, the results of both~\cite{GG-DCNN} and~\cite{Hou} are copied from~\cite{GG-DCNN}.  Note that~\cite{GG-DCNN} report results on downsampled images of size $256 \times 256$ and $512 \times 512$ (in line with their training settings). We reproduce the best scores of~\cite{GG-DCNN} and~\cite{Hou} in Table \ref{tab:MSCOCO_2K}. Also, since resizing to square images in this manner alters the aspect ratio, the results of our method as well as the remaining classical methods are reported on the original-sized images.
It can be seen from the results in Tables \ref{tab:UST_HK_1200}, \ref{tab:Kodak_data}, and \ref{tab:ESPL_v2_8_bit} that our method outperforms all competing algorithms on both animated and natural images across a diversity of bit depths and image resolutions, thereby validating the effectiveness of our bitplane-wise training strategy.

Qualitative results are shown in Fig. \ref{fig:qualitative}. Following prior literature~\cite{BitNet,BE-CNN,BDEN}, we use only images with a maximum bit depth of 8 for qualitative evaluation such that they are suitable for display on standard 8-bit monitors.
In the first example from the Kodak dataset~\cite{Kodak}, it can be observed that the vertical lines on the sails of the boat are best recovered by our method. In the second example from the ESPL v2 animated dataset~\cite{ESPL}, comparison methods either smooth the hair (e.g., CRR~\cite{CRR} and CA~\cite{CA}) or introduce artifacts reproducing the skin tones, while our result restores most of the fine details with hardly any noticeable artifacts. Additional results are provided in the supplementary.
\begin{figure}[!b]
%\begin{center}
\centering
\includegraphics[width=0.96\linewidth]{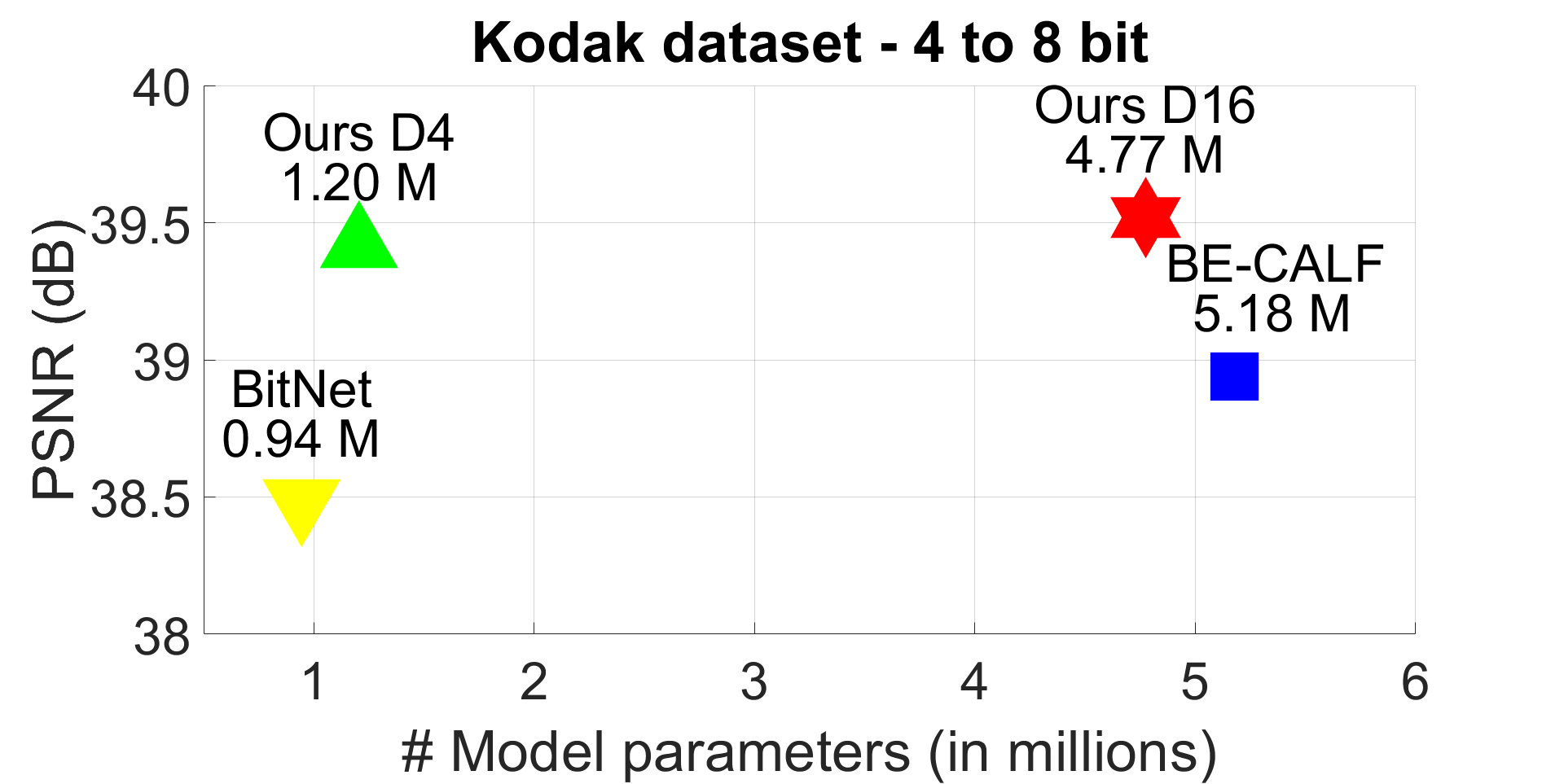}
%\end{center}
\caption{PSNR (dB) versus number of model parameters.
Our \# parameters is the total for all bitplanes.\label{fig:model_size}}
\end{figure}

\noindent \textbf{Accuracy versus model size:} The plot of Fig. \ref{fig:model_size} shows the PSNR (dB) versus number of model parameters on the Kodak dataset for the 4-8 bit scenario (see Table \ref{tab:Kodak_data}). Our method is compared against BE-CALF~\cite{BE-CALF} and BitNet~\cite{BitNet}, which are our two closest competitors. Our D4 model has approximately 1.2 million parameters in total for 4 bitplanes, while D16 requires 4.77 million. In comparison, BE-CALF, at 5.18 million parameters, is roughly four times bigger than D4 while yielding a PSNR about half a dB lower than our models. While BitNet has roughly the same model size as our D4 variant, its PSNR is roughly 1 dB lower than D4.

%\noindent \textbf{Single-shot baseline comparison:} Our proposed method employs $(N$-$q)$ networks with $D$ residual blocks each. For comparison, we trained a single network with the same total capacity by stacking together $(N$-$q)\times D$ residual blocks. The network was trained (under identical settings) to directly predict the residual using an MSE loss, with the final sigmoid layer removed. On the TESTIMAGES 1200 dataset~\cite{USTHK}, for example, for $D$=$4$, $N$=$8$, and $q$=$4$, the single-shot model obtained PSNR/SSIM values of 37.4822/0.9665, while our proposed method produced values of 39.6822/0.9691 (see Table~\ref{tab:UST_HK_1200}). This demonstrates the advantages of our bitplane-wise training strategy over the single-shot approach.

\noindent \textbf{Ablations on target and loss function:} To validate that our method's improvement accrues from our proposed bitplane-wise training strategy, we conducted two ablation studies. In the first ablation, we train $(N$-$q)$ networks to progressively improve image quality by predicting the next-bit quantized image $\mathbf{I}_{(q+k)}$, instead of the bitplane $\mathbf{B}_{N-(q+k)}$. The loss is changed to MSE since the target $\mathbf{I}_{(q+k)}$ is an image, and not a bitplane. All other training parameters are left unchanged. In Table~\ref{tab:new_ablations}, results are reported on four datasets for 4 to 8 bit recovery using the D4 model. Our proposed bitplane prediction, which is agnostic to the relative magnitude of the bit position, is more accurate than predicting the image $\mathbf{I}_{(q+k)}$.
We also performed a second ablation by changing only the loss function. In particular, we tested a perceptual VGG loss~\cite{Johnson2016Perceptual} and an MSE loss to predict the bitplane $\mathbf{B}_{N-(q+k)}$. Bitplanes have a very different distribution from natural images, and the VGG loss, which is intended to be applied to natural images, performs poorly. At the same time, the MSE loss offers competitive performance. This validates that the performance gain is due to our bitplane-wise training strategy, rather than the loss function. Our chosen BCE loss is more suitable since our target is a binary bitplane, and it offers slightly higher accuracy than MSE.

\begin{table}[t!]
\centering
\caption{Two ablation studies where we examine the effects of changing (i) the target, and (ii) the loss function. Results reported are for 4 to 8 bit recovery using the D4 model. Best results are in bold.}
\label{tab:new_ablations}
\scriptsize
\setlength{\tabcolsep}{2pt}
\begin{tabular}{C{2cm}C{1.25cm}C{1.5cm}C{1.5cm}C{1.25cm}}
\toprule
Dataset                & Sintel                                                   & MIT-Adobe                                & TESTIMAGES                                           & Kodak                                                    \\
 & \cite{Sintel} & FiveK~\cite{Adobe} & 1200~\cite{USTHK} & \cite{Kodak} \\ \toprule
Target -- Image           & \begin{tabular}[c]{@{}c@{}}37.4593\\ 0.9610\end{tabular} & \begin{tabular}[c]{@{}c@{}}37.3189\\ 0.9532\end{tabular} & \begin{tabular}[c]{@{}c@{}}35.8441\\ 0.9509\end{tabular} & \begin{tabular}[c]{@{}c@{}}37.2701\\ 0.9635\end{tabular} \\ \midrule
Loss -- VGG & \begin{tabular}[c]{@{}c@{}}35.4607\\ 0.9169\end{tabular} & \begin{tabular}[c]{@{}c@{}}34.7922\\ 0.9047\end{tabular} & \begin{tabular}[c]{@{}c@{}}35.2064\\ 0.9054\end{tabular} & \begin{tabular}[c]{@{}c@{}}34.2588\\ 0.9019\end{tabular} \\ \midrule
Loss -- MSE        & \begin{tabular}[c]{@{}c@{}}40.4113\\ 0.9767\end{tabular} & \begin{tabular}[c]{@{}c@{}}39.3000\\ 0.9660\end{tabular} & \begin{tabular}[c]{@{}c@{}}39.6185\\ 0.9684\end{tabular} & \begin{tabular}[c]{@{}c@{}}39.3259\\ 0.9702\end{tabular} \\ \midrule
Proposed               & \begin{tabular}[c]{@{}c@{}}\textbf{40.6143}\\ \textbf{0.9773}\end{tabular} & \begin{tabular}[c]{@{}c@{}}\textbf{39.4520}\\ \textbf{0.9668}\end{tabular} & \begin{tabular}[c]{@{}c@{}}\textbf{39.6822}\\ \textbf{0.9691}\end{tabular} & \begin{tabular}[c]{@{}c@{}}\textbf{39.4171}\\ \textbf{0.9709}\end{tabular} \\ \bottomrule
\end{tabular}%
\end{table}

\noindent \textbf{Accumulating bitplanes:} Table~\ref{tab:accum} shows that there is a performance gain as each lost bitplane is incrementally restored, and points to the effectiveness of bitplane-wise training. Since our method restores the image bitplane-by-bitplane, prediction errors can propagate. However, we found this effect to be minimal as also evident from our state-of-the-art results.
\begin{table}[t!]
\scriptsize
\caption{TESTIMAGES 1200 dataset~\cite{USTHK} for 4 to 8-bit recovery, with $D$=$16$. To compute metrics, we use as ground truth the image quantized upto the corresponding bit depth. Performance improves as each lost bitplane (from 5 to 8) is recovered.
\label{tab:accum}
}
\begin{center}
\setlength{\tabcolsep}{4pt}
\begin{tabular}{ccccc}
%\centering
\toprule
Bitplane accumulated & 5th & 6th & 7th & 8th\\
\toprule
PSNR (dB) & 38.5700 & 39.4639 & 40.1529 & 40.3906\\
SSIM &  0.9509 & 0.9602 & 0.9690 & 0.9725\\
\bottomrule
\end{tabular}
\end{center}
\end{table}

\noindent \textbf{Running time:} On an Nvidia Tesla V100
GPU with 32 GB of RAM, our D4 model takes approximately 0.08 seconds per bitplane to process a $768 \times 512$ resolution image, while our D16 model takes 0.28 seconds on average.
The running times of competing methods are reported in the supplementary material.

\section{Conclusion}
We have proposed a new DNN training strategy for bit-depth recovery where the residual image corresponding to the bits lost due to quantization is recovered bitplane by bitplane. Bitplane-wise learning makes our method independent of the relative magnitude of the bit position, overcoming the limitation of existing single-shot approaches. Experiments on five benchmark datasets demonstrate that our method outperforms the state of the art by up to 2.3 dB PSNR depending on the quantization level. As future work, we plan to explore the sharing of features across bitplanes instead of training networks completely independent of each other. We also plan to examine bitplane-wise training for other residual estimation problems.

\newcommand{\hbAppendixPrefix}{S}
\renewcommand{\thefigure}{\hbAppendixPrefix\arabic{figure}}
\setcounter{figure}{0}
\renewcommand{\thetable}{\hbAppendixPrefix\arabic{table}}
\setcounter{table}{0}
\renewcommand{\theequation}{\hbAppendixPrefix\arabic{equation}}
\setcounter{equation}{0}
\renewcommand{\thesection}{\hbAppendixPrefix\arabic{section}}
\setcounter{section}{0}

\clearpage

\twocolumn[{%
 \centering
 \Huge Supplementary Material\\[1.5em]
}]

%\title{Supplementary Material \\ A Little Bit More:\\Bitplane-Wise Bit-Depth Recovery}
%
%
%\newpage
%
%\maketitle

This supplementary material provides additional results and details that could not be included in the main paper due to space constraints. Ablation studies on our network architecture are presented in Sec.~\ref{sec:ablation}. A single-shot baseline and a comparison against the standard U-Net~\cite{UNet} architecture are reported in Sec.~\ref{sec:baselines}. We also perform experiments to validate our training protocol in Sec.~\ref{sec:joint_training}, and choice of training data in Sec.~\ref{sec:training_data}. In Secs.~\ref{sec:quantitative} and~\ref{sec:qualitative}, we provide further quantitative and qualitative comparisons and results. The running times of various methods are reported in Sec.~\ref{sec:time}. An example highlighting the importance of bit-depth recovery for photo editing is furnished in Sec.~\ref{sec:editing}.

\section{Ablations on network architecture}% for different $D$ values}
\begin{figure}[b!]
\begin{center}
\includegraphics[width=0.95\linewidth]{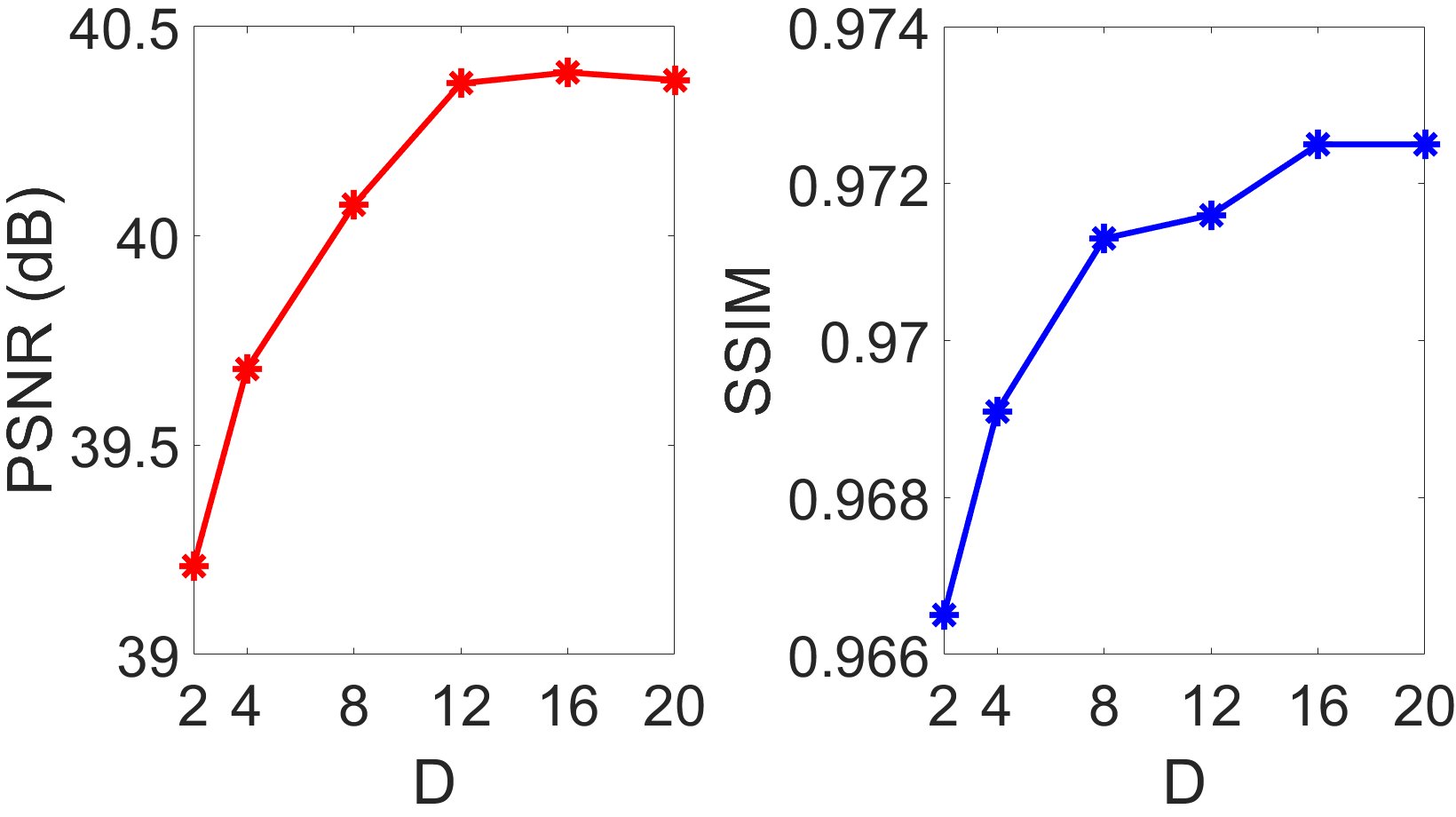}
\end{center}
\caption{PSNR (dB) and SSIM versus number of residual units $D$ in our network architecture on the TESTIMAGES 1200 dataset~\cite{USTHK}. Results reported are for 4 to 8 bit recovery.
\label{fig:ablation}}
\end{figure}

\label{sec:ablation}
We performed ablation studies by varying the number of residual units $D$ in our network architecture (see Fig. 4 of our main paper). Specifically, we tested the following values $D = 2,4,8,12,16,20$. The average PSNR (dB) and SSIM versus $D$ on the TESTIMAGES 1200 dataset~\cite{USTHK} for 4 to 8 bit recovery are shown in the plots of Fig. \ref{fig:ablation}. We have reported results, both in the main paper and this supplementary, for our D4 and D16 variants. We have showed that we already outperform existing methods with $D=4$. Beyond $D=16$, we did not observe an improvement in performance.
\begin{figure}[t!]
\begin{center}
\includegraphics[width=0.975\linewidth]{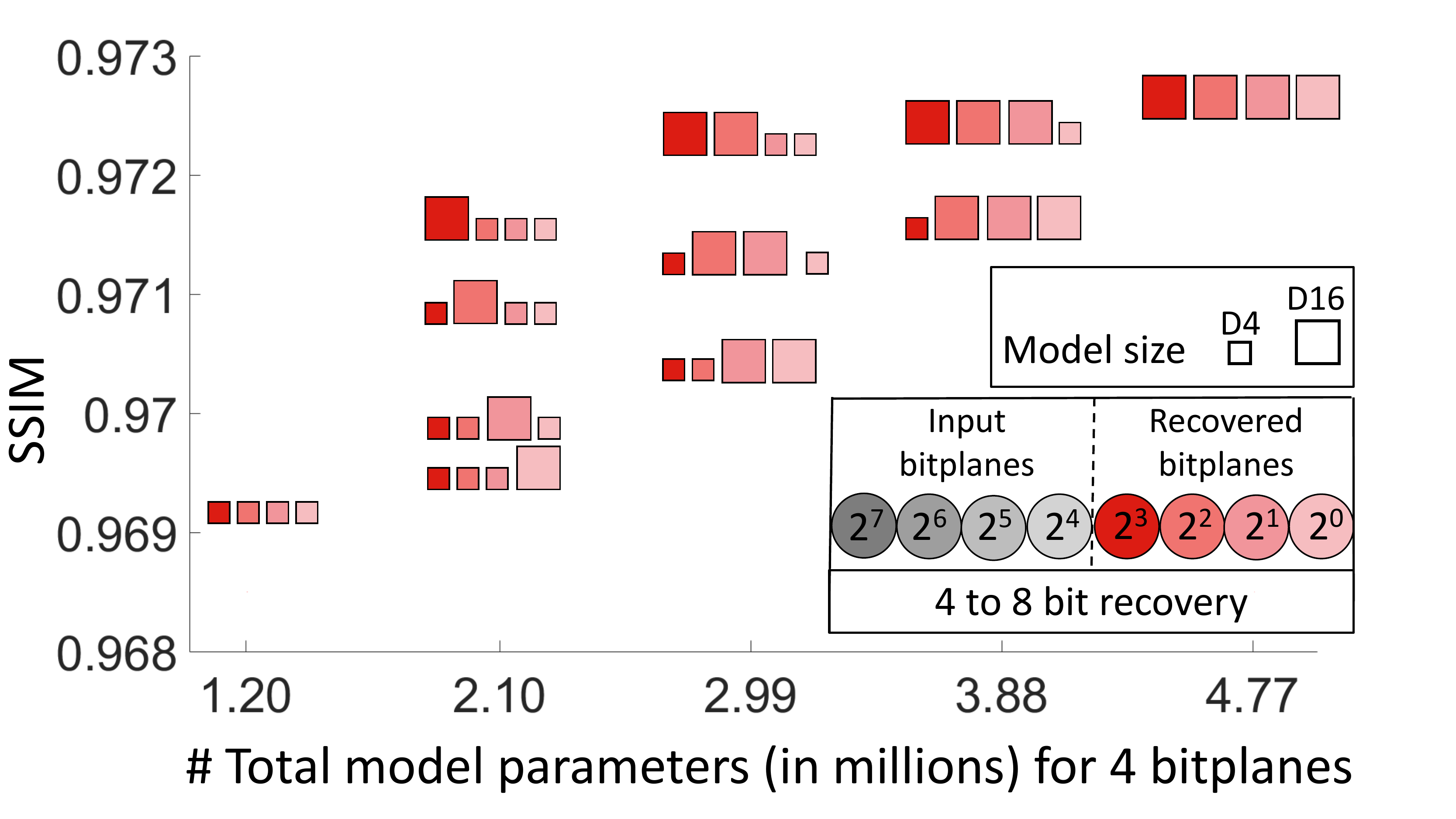}
\end{center}
\caption{An ablation study where the network architecture is varied for each bitplane. We experimented with different permutations of our D4 and D16 models to recover the bitplanes. The plot shows SSIM versus total number of network parameters (in millions) for 4 to 8 bit recovery on the TESTIMAGES 1200 dataset~\cite{USTHK}. The number of D16 models increases from zero to four from left to right.
\label{fig:ablation2}}
\end{figure}

We also performed an ablation study where we varied the network architecture for each bitplane. In particular, we tested the 4 to 8 bit recovery scenario with each bitplane being recovered by either our D4 model or our D16 model. The plot of Fig. \ref{fig:ablation2} shows SSIM versus the total number of network parameters (in millions) on the TESTIMAGES 1200 dataset~\cite{USTHK}. In the first column with 1.20 million paramters, the D4 model was used to recover all four bitplanes. The number of D16 models used for recovery increases by one in each subsequent column, with the last column using D16 models for all four bitplanes. It can be observed from the plot that the most gains are achieved by using the larger D16 model for the more significant bit positions e.g., $2^3$ and $2^2$ in the present case.

\begin{table*}[h]
\centering
\caption{An ablation study where we replace our network architecture with a standard U-Net~\cite{UNet}. Two different U-Nets were trained with model sizes approximately equivalent to our D4 and D16 models. Results are presented for 4 to 8 bit recovery. Best results are in bold.}
\label{tab:R13}
%\begin{tabular}{|c|c|c|c|c|c|c|c|c|}
\setlength{\tabcolsep}{1pt}
\begin{tabular}{C{2cm}C{1.5cm}C{1.5cm}C{1.5cm}C{1.5cm}C{1.5cm}C{1.5cm}C{1.5cm}C{1.5cm}C{1.5cm}}
\toprule
Dataset    & & \multicolumn{2}{c}{Sintel~\cite{Sintel}}                                                                                         & \multicolumn{2}{c}{TESTIMAGES 1200~\cite{USTHK}}                                                                                & \multicolumn{2}{c}{Kodak~\cite{Kodak}}                                                                                          & \multicolumn{2}{c}{ESPL v2~\cite{ESPL}}                                                                                        \\ \toprule
Model size & & D4                                                       & D16                                                      & D4                                                       & D16                                                      & D4                                                       & D16                                                      & D4                                                       & D16                                                      \\ \midrule
U-Net    &  \begin{tabular}[c]{@{}c@{}}PSNR\\ SSIM\end{tabular}  & \begin{tabular}[c]{@{}c@{}}39.4725\\ 0.9699\end{tabular} & \begin{tabular}[c]{@{}c@{}}40.2996\\ 0.9751\end{tabular} & \begin{tabular}[c]{@{}c@{}}38.5855\\ 0.9604\end{tabular} & \begin{tabular}[c]{@{}c@{}}39.2869\\ 0.9657\end{tabular} & \begin{tabular}[c]{@{}c@{}}38.5757\\ 0.9690\end{tabular} & \begin{tabular}[c]{@{}c@{}}38.9416\\ 0.9691\end{tabular} & \begin{tabular}[c]{@{}c@{}}38.8132\\ 0.9507\end{tabular} & \begin{tabular}[c]{@{}c@{}}39.2789\\ 0.9526\end{tabular} \\ \midrule
Ours    & \begin{tabular}[c]{@{}c@{}}PSNR\\ SSIM\end{tabular} &    \begin{tabular}[c]{@{}c@{}}\textbf{40.6143}\\ \textbf{0.9773}\end{tabular} & \begin{tabular}[c]{@{}c@{}}\textbf{41.1909}\\ \textbf{0.9794}\end{tabular} & \begin{tabular}[c]{@{}c@{}}\textbf{39.6822}\\ \textbf{0.9691}\end{tabular} & \begin{tabular}[c]{@{}c@{}}\textbf{40.3906}\\ \textbf{0.9725}\end{tabular} & \begin{tabular}[c]{@{}c@{}}\textbf{39.4171}\\ \textbf{0.9709}\end{tabular} & \begin{tabular}[c]{@{}c@{}}\textbf{39.5185}\\ \textbf{0.9723}\end{tabular} & \begin{tabular}[c]{@{}c@{}}\textbf{39.3854}\\ \textbf{0.9532}\end{tabular} & \begin{tabular}[c]{@{}c@{}}\textbf{39.5312}\\ \textbf{0.9528}\end{tabular} \\ \bottomrule
\end{tabular}
\end{table*}

\begin{table*}[h]
\centering
\caption{An ablation study where the bitplane networks are jointly trained in an end-to-end manner, instead of independently. Results are presented for 4 to 8, 4 to 16, and 4 to 12 bit recovery using our D4 model. Best results are in bold.}
\label{tab:R12}
\setlength{\tabcolsep}{1pt}
\begin{tabular}{C{3cm}C{1.5cm}C{2cm}C{2cm}C{2.5cm}C{2cm}C{2cm}C{2cm}} \toprule
 \multicolumn{8}{c}{4 to 8 bit recovery} \\
 \toprule
\multirow{2}{*}{Dataset}     &  & Sintel                                                   & MIT-Adobe                                                                         & TESTIMAGES                                          & Kodak                                                    & MSCOCO                                                   & BSD100                                                   \\
& & \cite{Sintel} & FiveK~\cite{Adobe} & 1200~\cite{USTHK} & \cite{Kodak} & \cite{MSCOCO} & \cite{BSD} \\
 \toprule
\begin{tabular}[c]{@{}c@{}}Jointly trained\\ for 4 to 8 bit recovery \end{tabular}   & \begin{tabular}[c]{@{}c@{}}PSNR\\ SSIM\end{tabular} & \begin{tabular}[c]{@{}c@{}}40.1737\\ 0.9754\end{tabular} & \multicolumn{1}{c}{\begin{tabular}[c]{@{}c@{}}39.4463\\ \textbf{0.9681}\end{tabular}} & \begin{tabular}[c]{@{}c@{}}\textbf{39.7444}\\ 0.9688\end{tabular} & \begin{tabular}[c]{@{}c@{}}\textbf{39.4597}\\ \textbf{0.9716}\end{tabular} & \begin{tabular}[c]{@{}c@{}}\textbf{39.4223}\\ 0.9615\end{tabular} & \begin{tabular}[c]{@{}c@{}}39.1204\\ \textbf{0.9721}\end{tabular} \\ \midrule
\begin{tabular}[c]{@{}c@{}}Independently trained\\ Proposed\end{tabular}       & \begin{tabular}[c]{@{}c@{}}PSNR\\ SSIM\end{tabular} & \begin{tabular}[c]{@{}c@{}}\textbf{40.6143}\\ \textbf{0.9773}\end{tabular} & \begin{tabular}[c]{@{}c@{}}\textbf{39.4520}\\ 0.9668\end{tabular}                      & \begin{tabular}[c]{@{}c@{}}39.6822\\ \textbf{0.9691}\end{tabular} & \begin{tabular}[c]{@{}c@{}}39.4171\\ 0.9709\end{tabular} & \begin{tabular}[c]{@{}c@{}}39.4171\\ \textbf{0.9625}\end{tabular} & \begin{tabular}[c]{@{}c@{}}\textbf{39.1499}\\ 0.9709\end{tabular} \\ 

%\bottomrule
\end{tabular}
\begin{tabular}{C{3cm}C{1.5cm}C{2cm}C{2cm}C{2.5cm}|C{2cm}C{2cm}C{2cm}} 
 \toprule
\multicolumn{2}{c}{} & \multicolumn{3}{c|}{4 to 16 bit recovery} & \multicolumn{3}{c}{4 to 12 bit recovery}\\
\toprule 
\multirow{2}{*}{Dataset}     &  & Sintel                                                   & MIT-Adobe                                                                         & TESTIMAGES                                          & Sintel & MIT-Adobe                                                   & TESTIMAGES                                                   \\
& & \cite{Sintel} & FiveK~\cite{Adobe} & 1200~\cite{USTHK} & \cite{Sintel} & FiveK~\cite{Adobe} & 1200~\cite{USTHK} \\
 \toprule

\begin{tabular}[c]{@{}c@{}}Jointly trained\\ for 4 to 16 bit recovery \end{tabular}                                                          & \multicolumn{1}{c}{\begin{tabular}[c]{@{}c@{}}PSNR\\ SSIM\end{tabular}} & \begin{tabular}[c]{@{}c@{}}40.6079\\ 0.9774\end{tabular} & \multicolumn{1}{c}{\begin{tabular}[c]{@{}c@{}}\textbf{39.7671}\\ \textbf{0.9703}\end{tabular}} & \begin{tabular}[c]{@{}c@{}}\textbf{39.8270}\\ \textbf{0.9700}\end{tabular} & \begin{tabular}[c]{@{}c@{}}40.6074\\ 0.9774\end{tabular}                                                                                                               & \begin{tabular}[c]{@{}c@{}}\textbf{39.7633}\\ \textbf{0.9702}\end{tabular}                                                                                               & \begin{tabular}[c]{@{}c@{}}\textbf{39.8350}\\ \textbf{0.9700}\end{tabular}                                                                                                             \\ \midrule
\begin{tabular}[c]{@{}c@{}}Independently trained\\ Proposed\end{tabular} & \multicolumn{1}{c}{\begin{tabular}[c]{@{}c@{}}PSNR\\ SSIM\end{tabular}} & \begin{tabular}[c]{@{}c@{}}\textbf{40.9274}\\ \textbf{0.9786}\end{tabular} & \begin{tabular}[c]{@{}c@{}}39.6484\\ 0.9683\end{tabular}                      & \begin{tabular}[c]{@{}c@{}}39.6503\\ \textbf{0.9700}\end{tabular} & \begin{tabular}[c]{@{}c@{}}\textbf{40.9286}\\ \textbf{0.9786}\end{tabular}                                                        & \begin{tabular}[c]{@{}c@{}}39.6467\\ 0.9683\end{tabular}                                                                 & \begin{tabular}[c]{@{}c@{}}39.6619\\ \textbf{0.9700}\end{tabular}                                                                                 \\ \bottomrule
\end{tabular}

\end{table*}

\section{Baseline comparisons}
\label{sec:baselines}
We perform two baseline comparisons. The first is a single-shot approach based on our proposed network architecture, while the second is our proposed multi-level training strategy but with our model replaced by a standard U-Net~\cite{UNet}.

\noindent \textbf{Single-shot baseline comparison:} Our proposed method employs $(N$-$q)$ networks with $D$ residual blocks each. For comparison, we trained a single network with the same total capacity by stacking together $(N$-$q)\times D$ residual blocks. The network was trained (under identical settings) to directly predict the residual using an MSE loss, with the final sigmoid layer removed. On the TESTIMAGES 1200 dataset~\cite{USTHK}, for example, for $D$=$4$, $N$=$8$, and $q$=$4$, the single-shot model obtained PSNR/SSIM values of 37.4822/0.9665, while our proposed method produced values of 39.6822/0.9691. This demonstrates the advantages of our bitplane-wise training strategy over the single-shot approach.

\noindent \textbf{Comparison against U-Net~\cite{UNet}:} Our proposed bitplane-wise training scheme is a general strategy that can be implemented using any network architecture of choice. In Table~\ref{tab:R13}, we compare against the standard U-Net~\cite{UNet} model for 4 to 8 bit recovery. We varied the number of filters in the initial level of the U-Net (the number of filters in the subsequent layers are multiples thereof) such that the total number of parameters in the model is approximately equal to our proposed network. In particular, our D4 model has 301,248 parameters per bitplane while our D16 model has 1,193,664 parameters. For a fair comparison, we trained two U-Net models with 7 and 13 filters in the initial level for a total of 372,088 parameters (D4 equivalent) and 1,281,790 parameters (D16 equivalent) per bitplane, respectively. In can be observed from Table~\ref{tab:R13} that the U-Net models also offer competitive performance, but our proposed architecture is more accurate.\\

\begin{table*}[h]
\centering
\caption{An ablation study where we train on either natural images or animated images, instead of a mixture of the two. Results are presented for 4 to 8 bit recovery using our D4 model. Best results are in bold.}
\label{tab:R232}
\setlength{\tabcolsep}{1pt}
\begin{tabular}{C{3cm}C{1.5cm}C{2cm}C{2cm}C{2.5cm}C{2cm}C{2cm}C{2cm}} \toprule
\multirow{2}{*}{Dataset} &                                                             & Sintel                                                   & MIT-Adobe                               & TESTIMAGES                                           & Kodak                                                    & ESPL v2                                                  & BSD100                                                   \\
& & \cite{Sintel} & FiveK~\cite{Adobe}  & 1200~\cite{USTHK} & \cite{Kodak} & \cite{ESPL} & \cite{BSD} \\ \toprule
Natural only                                                       &  \begin{tabular}[c]{@{}c@{}}PSNR\\ SSIM\end{tabular} & \begin{tabular}[c]{@{}c@{}}39.9516\\ 0.9755\end{tabular} & \begin{tabular}[c]{@{}c@{}}\textbf{39.5514}\\ \textbf{0.9669}\end{tabular} & \begin{tabular}[c]{@{}c@{}}\textbf{39.9929}\\ \textbf{0.9699}\end{tabular} & \begin{tabular}[c]{@{}c@{}}39.3899\\ 0.9702\end{tabular} & \begin{tabular}[c]{@{}c@{}} 38.9856\\ 0.9502\end{tabular} & \begin{tabular}[c]{@{}c@{}}39.1344\\ 0.9690\end{tabular} \\ \midrule
Animated only                                                      &  \begin{tabular}[c]{@{}c@{}}PSNR\\ SSIM\end{tabular} & \begin{tabular}[c]{@{}c@{}}\textbf{40.7571}\\ \textbf{0.9778}\end{tabular} & \begin{tabular}[c]{@{}c@{}}38.6983\\ 0.9632\end{tabular} & \begin{tabular}[c]{@{}c@{}}38.6808\\ 0.9641\end{tabular} & \begin{tabular}[c]{@{}c@{}}39.3094\\ 0.9707\end{tabular} & \begin{tabular}[c]{@{}c@{}}39.3593\\ 0.9512 \end{tabular} & \begin{tabular}[c]{@{}c@{}}38.9220\\ 0.9704\end{tabular} \\ \midrule
\begin{tabular}[c]{@{}c@{}} Natural+animated \\ Proposed\end{tabular} &  \begin{tabular}[c]{@{}c@{}}PSNR\\ SSIM\end{tabular} & \begin{tabular}[c]{@{}c@{}}40.6143\\ 0.9773\end{tabular} & \begin{tabular}[c]{@{}c@{}}39.4520\\ 0.9668\end{tabular} & \begin{tabular}[c]{@{}c@{}}39.6822\\ 0.9691\end{tabular} & \begin{tabular}[c]{@{}c@{}}\textbf{39.4171}\\ \textbf{0.9709}\end{tabular} & \begin{tabular}[c]{@{}c@{}}\textbf{39.3854}\\ \textbf{0.9532}\end{tabular} & \begin{tabular}[c]{@{}c@{}}\textbf{39.1499}\\ \textbf{0.9709}\end{tabular} \\ \bottomrule
\end{tabular}
\end{table*}

\section{Training protocol}
\label{sec:joint_training}

In our proposed framework, the $(N$-$q)$ bitplane networks are trained independently. This means that from the second network onwards, the distribution of the input $\mathbf{\widehat{I}}_{(q+k-1)}$ at test time is not identical to the distribution of the input $\mathbf{I}_{(q+k-1)}$ during training. To verify whether this difference in distribution has a significant impact on accuracy, we jointly trained all $(N$-$q)$ networks in an end-to-end fashion, and compared performance against our proposed independently trained networks. It is important to note that our proposed approach has the advantage that the network corresponding to each bitplane needs to be trained only once, whereas for joint end-to-end training, these networks have to be re-trained for each input quantization. For example, the network that predicts the 5$^{\mbox{\scriptsize{th}}}$ bitplane trained for 4 to 8 bit recovery cannot be re-used for 3 to 8 bit recovery under the joint training pipeline.

The results of joint training are presented in Table~\ref{tab:R12}. In all cases, the D4 model was used. We first tested our method against a model trained jointly for 4 to 8 bit recovery. We also trained another joint model for 4 to 16 bit recovery, and tested it on the 4 to 16 and 4 to 12 bit recovery tasks. It can be observed that joint training is more accurate in some scenarios, while our proposed method works better in other cases. However, the difference in performance is very small. Overall, we did not find that joint training offers an improvement that outweighs the benefits of independent training, wherein each bitplane network needs to be trained only once.

\section{Choice of training data}
\label{sec:training_data}
Our proposed models were trained on 2000 images, 1000 natural images from the MIT-Adobe FiveK dataset~\cite{Adobe} and 1000 animated images from the Sintel dataset~\cite{Sintel}. We performed an ablation experiment where we trained two separate models -- (1) on 2000 natural images from the MIT-Adobe FiveK dataset, and (2) on 2000 animated images from the Sintel dataset. The results on six different test sets are presented in Table~\ref{tab:R232}. Scores reported are using our D4 model for 4 to 8 bit recovery. As expected, training on only animated images gives the best performance on the Sintel animation dataset, while training only on natural images produces the best accuracy on the MIT-Adobe FiveK natural image dataset. However, there is a 0.81 dB drop in PSNR when the model trained on natural images is tested on animated images, and similarly, a 0.85 dB drop when the model trained on animated images is tested on natural images. In comparison, our mixed model performs well on both natural and animated images with only a nominal drop of 0.14 dB and 0.10 dB, respectively, compared to the best model. Similar observations can be made on the other test sets. TESTIMAGES 1200~\cite{USTHK}, Kodak~\cite{Kodak} and BSD100~\cite{BSD} are natural image datasets, and the model trained on only animated images incurs a significant drop in performance, while our mixed model performs almost on par or even better than the model trained purely on natural images. Likewise, on the ESPL v2~\cite{ESPL} animated dataset, the model trained purely on natural images suffers a drop in performance, while our mixed model is the most accurate.

\section{Additional quantitative results}
\label{sec:quantitative}
We compare against the deep learning methods of Hou et al.~\cite{Hou} and GG-DCNN~\cite{GG-DCNN}, who have trained and tested their models on the Microsoft COCO~\cite{MSCOCO} dataset. Their pre-trained models/codes are not available. Following GG-DCNN~\cite{GG-DCNN}, we use 2000 random images from the testing fold of this dataset. The authors of ~\cite{GG-DCNN} have re-implemented the method of~\cite{Hou}, and in Table \ref{tab:MSCOCO_2K}, the results of both~\cite{GG-DCNN} and~\cite{Hou} are copied from~\cite{GG-DCNN}.  Note that~\cite{GG-DCNN} report results on downsampled images of size $256 \times 256$ and $512 \times 512$ (in line with their training settings). We reproduce the best scores of~\cite{GG-DCNN} and~\cite{Hou} in Table \ref{tab:MSCOCO_2K}. Also, since resizing to square images in this manner alters the aspect ratio, the results of our method are reported on the original-sized images. It can be observed from the results in Table \ref{tab:MSCOCO_2K} that our PSNR values are significantly higher than~\cite{Hou} and~\cite{GG-DCNN}.  

\begin{table}[]
\caption{Results on Microsoft COCO dataset~\cite{MSCOCO}. Following the convention used in the main paper, the best results are reported in bold and red. The second, third, and fourth best-performing methods are shown in green, blue, and yellow, respectively. NR denotes that a score was `not reported' in the original paper. Numbers copied directly from the original papers are marked with a $\dagger$ symbol.
\label{tab:MSCOCO_2K}}
\begin{center}
\scriptsize
\setlength{\tabcolsep}{2pt}
%\begin{tabular}{cc>{\centering\arraybackslash}m{1.4cm}>{\centering\arraybackslash}m{1.66cm}>{\centering\arraybackslash}m{1.4cm}>{\centering\arraybackslash}m{1.4cm}}
\begin{tabular}{C{0.65cm}C{0.65cm}C{1cm}C{1.76cm}C{1.3cm}C{1.1cm}C{1.1cm}}
\toprule
%\multicolumn{2}{c}{} &  Hou & GG-DCNN & Ours D4 & Ours D16 \\
%\multicolumn{2}{c}{} &  \cite{Hou} & \cite{GG-DCNN} &  & \\
\multicolumn{2}{c}{} &  Hou~\cite{Hou} & GG-DCNN \cite{GG-DCNN} & BitNet~\cite{BitNet} & Ours D4 & Ours D16 \\
\toprule
\multirow{2}{*}{4-8 bit} & PSNR &  35.8800$^{\dagger}$ & \cellcolor{yellow!75} 37.5300$^{\dagger}$ & \cellcolor{blue!25} 37.9678 & \cellcolor{green!25} 39.4171 & \cellcolor{red!25} \textbf{39.4344} \\

 	 	 	 	 	 	  & SSIM & NR & NR & \cellcolor{blue!25} 0.9529 & \cellcolor{green!25} 0.9625 & \cellcolor{red!25} \textbf{0.9629} \\
\bottomrule
\end{tabular}
\end{center}
\end{table}

\begin{table*}[]
\caption{Results on TESTIMAGES 800 dataset~\cite{USTHK}.
\label{tab:UST_HK_800}}
\begin{center}
\scriptsize
\setlength{\tabcolsep}{4pt}
\begin{tabular}{ccccccccccccc}
\toprule
\multicolumn{2}{c}{} & ZP & MIG & BR & MRC & CRR & CA & ACDC & IPAD & BitNet & Ours D4 & Ours D16 \\
\multicolumn{2}{c}{} & & & \cite{BR} & \cite{MRC} & \cite{CRR} & \cite{CA} & \cite{ACDC} & \cite{IPAD}  & \cite{BitNet} &  & \\
\toprule
\multirow{2}{*}{3-16 bit} & PSNR & 22.7573 & 25.2648 & 26.4243 & 27.5546 & 26.4007 & 28.8859 & 28.8005 & \cellcolor{yellow!75} 29.4943 & \cellcolor{blue!25} 32.1362 & \cellcolor{green!25} 32.9752 & \cellcolor{red!25} \textbf{33.4297} \\ 

 	 	 	 	 	 	  & SSIM & 0.7550 & 0.7508 & 0.7965 & 0.8212 & 0.8393 & 0.8612 & 0.8199 & \cellcolor{yellow!75} 0.8739 & \cellcolor{blue!25} 0.9052 & \cellcolor{green!25} 0.9237 & \cellcolor{red!25} \textbf{0.9291} \\ 
\midrule
\multirow{2}{*}{4-16 bit} & PSNR & 28.8342 & 31.5370 & 32.0966 & 34.3637 & 33.2068 & 34.8206 & 34.7631 & \cellcolor{yellow!75} 35.5728 & \cellcolor{blue!25} 38.3376 & \cellcolor{green!25} 39.3081 & \cellcolor{red!25} \textbf{40.0830} \\ 

 	 	 	 	 	 	  & SSIM & 0.8846 & 0.8822 & 0.8946 & 0.9233 & 0.9237 & 0.9336 & 0.9086 & \cellcolor{yellow!75} 0.9417 & \cellcolor{blue!25} 0.9587 & \cellcolor{green!25} 0.9699 & \cellcolor{red!25} \textbf{0.9734} \\ 
\midrule
\multirow{2}{*}{5-16 bit} & PSNR & 34.8447 & 37.6917 & 37.9665 & 40.7613 & 39.3272 & 40.1733 & 40.7694 & \cellcolor{yellow!75} 41.1885 & \cellcolor{blue!25} 43.7923 & \cellcolor{green!25} 45.1050 & \cellcolor{red!25} \textbf{45.8028} \\ 

 	 	 	 	 	 	  & SSIM & 0.9574 & 0.9571 & 0.9597 & 0.9727 & 0.9670 & 0.9717 & 0.9649 & \cellcolor{yellow!75} 0.9749 & \cellcolor{blue!25} 0.9840 & \cellcolor{green!25} 0.9890 & \cellcolor{red!25} \textbf{0.9903} \\ 
\midrule
\multirow{2}{*}{6-16 bit} & PSNR & 40.8653 & 43.8044 & 43.9379 & \cellcolor{yellow!75} 46.9165 & 45.0570 & 45.0097 & 46.7881 & 46.6101 & \cellcolor{blue!25} 48.6751 & \cellcolor{green!25} 50.6359 & \cellcolor{red!25} \textbf{51.0936} \\ 

 	 	 	 	 	 	  & SSIM & 0.9872 & 0.9873 & 0.9877 & \cellcolor{yellow!75} 0.9918 & 0.9863 & 0.9887 & 0.9889 & 0.9902 & \cellcolor{blue!25} 0.9943 & \cellcolor{green!25} 0.9964 & \cellcolor{red!25} \textbf{0.9966} \\ 
\bottomrule
\end{tabular}
\end{center}
\end{table*}

The TESTIMAGES dataset~\cite{USTHK} contains 40 16-bit natural images. While~\cite{BE-CALF,BE-CNN} have reported results on the $1200 \times 1200$ resolution variant of this dataset, BitNet~\cite{BitNet} has used $800 \times 800$ resolution images in their paper. We reported comparisons in Table 3 of our main paper on $1200 \times 1200$ resolution images. In Table~\ref{tab:UST_HK_800} of this supplementary, we present comparisons on $800 \times 800$ resolution images. (Following~\cite{BitNet}, we used image files with the shifting indicator `B01C00'.) From the results, it can be observed that both our models, D4 and D16, outperform BitNet on both metrics.

\begin{table*}[]
\caption{8--10 bit and 8--12 bit recovery.
\label{tab:8_10_12}}
\begin{center}
\scriptsize
\setlength{\tabcolsep}{4pt}
\begin{tabular}{ccccccccccccccc}
\toprule
\multicolumn{2}{c}{} & ZP & MIG & BR & MRC & CRR & CA & ACDC & IPAD & BE-CNN & BE-CALF & BitNet & Ours D4 & Ours D16 \\
\multicolumn{2}{c}{} & & & \cite{BR} & \cite{MRC} & \cite{CRR} & \cite{CA} & \cite{ACDC} & \cite{IPAD}  & \cite{BE-CNN} & \cite{BE-CALF} & \cite{BitNet} &  & \\
\toprule
\multicolumn{15}{c}{\small {Results on Sintel dataset~\cite{Sintel}}}\\
\toprule
\multirow{2}{*}{8-10 bit} & PSNR & 54.6856 & 57.2202 & 56.3490 & \cellcolor{blue!25} 59.5132 & 58.3145 & 58.2976 & 58.2234 & 58.4633 & 54.7227 & \cellcolor{yellow!75} 59.2435 & 56.9947 & \cellcolor{green!25} 60.8031 & \cellcolor{red!25} \textbf{60.9087} \\ 

 	 	 	 	 	 	  & SSIM & 0.9991 & 0.9991 & 0.9991 & 0.9993 & 0.9991 & 0.9992 & 0.9989 & \cellcolor{yellow!75} 0.9993 & 0.9989 & \cellcolor{blue!25} 0.9993 & 0.9990 & \cellcolor{green!25} 0.9997 & \cellcolor{red!25} \textbf{0.9997} \\ 
\midrule
\multirow{2}{*}{8-12 bit} & PSNR & 53.2583 & 56.6246 & 56.4774 & \cellcolor{yellow!75} 59.4131 & 57.7446 & 58.1010 & 58.6677 & 58.7444 & 54.7821 & \cellcolor{blue!25} 59.5138 & 57.4428 & \cellcolor{green!25} 62.9363 & \cellcolor{red!25} \textbf{63.2755} \\ 

 	 	 	 	 	 	  & SSIM & 0.9990 & 0.9990 & 0.9990 & \cellcolor{yellow!75} 0.9993 & 0.9984 & 0.9989 & 0.9989 & 0.9990 & 0.9989 & \cellcolor{blue!25} 0.9993 & 0.9989 & \cellcolor{green!25} 0.9997 & \cellcolor{red!25} \textbf{0.9998} \\ 
\toprule
\multicolumn{15}{c}{\small {Results on TESTIMAGES 1200 dataset~\cite{USTHK}}}\\
\toprule
\multirow{2}{*}{8-10 bit} & PSNR & 54.7353 & 56.8199 & 55.5634 & \cellcolor{blue!25} 59.3286 & 57.9548 & 55.9924 & \cellcolor{yellow!75} 58.3119 & 57.9176 & 53.0972 & 58.0078 & 53.2962 & \cellcolor{green!25} 59.9239 & \cellcolor{red!25} \textbf{60.0357} \\ 

 	 	 	 	 	 	  & SSIM & 0.9992 & 0.9991 & 0.9991 & \cellcolor{yellow!75} 0.9993 & 0.9991 & 0.9991 & 0.9991 & 0.9992 & 0.9986 & \cellcolor{blue!25} 0.9993 & 0.9971 & \cellcolor{green!25} 0.9996 & \cellcolor{red!25} \textbf{0.9996} \\ 
\midrule
\multirow{2}{*}{8-12 bit} & PSNR & 53.3181 & 56.1645 & 55.9193 & \cellcolor{blue!25} 59.1269 & 57.2560 & 55.6482 & \cellcolor{yellow!75} 58.7748 & 57.9785 & 53.1392 & 58.1207 & 53.5826 & \cellcolor{green!25} 61.2724 & \cellcolor{red!25} \textbf{61.5413} \\ 

 	 	 	 	 	 	  & SSIM & 0.9990 & 0.9990 & 0.9990 & \cellcolor{yellow!75} 0.9993 & 0.9985 & 0.9988 & 0.9991 & 0.9989 & 0.9986 & \cellcolor{blue!25} 0.9993 & 0.9971 & \cellcolor{green!25} 0.9996 & \cellcolor{red!25} \textbf{0.9996} \\ 
\toprule
\multicolumn{15}{c}{\small {Results on TESTIMAGES 800 dataset~\cite{USTHK}}}\\
\toprule
\multirow{2}{*}{8-10 bit} & PSNR & 54.7371 & 56.8124 & 55.5558 & \cellcolor{blue!25} 59.1069 & 57.9243 & 56.6157 & \cellcolor{yellow!75} 58.3242 & 57.7774 & 51.0963 & 57.5190 & 52.9422 & \cellcolor{green!25} 59.3925 & \cellcolor{red!25} \textbf{59.4411} \\ 

 	 	 	 	 	 	  & SSIM & 0.9993 & 0.9993 & 0.9992 & \cellcolor{blue!25} 0.9994 & 0.9992 & 0.9992 & 0.9992 & 0.9993 & 0.9982 & \cellcolor{yellow!75} 0.9993 & 0.9973 & \cellcolor{green!25} 0.9996 & \cellcolor{red!25} \textbf{0.9996} \\ 
\midrule
\multirow{2}{*}{8-12 bit} & PSNR & 53.3218 & 56.1560 & 55.9105 & \cellcolor{blue!25} 58.9844 & 57.2802 & 56.2646 & \cellcolor{yellow!75} 58.7920 & 57.8165 & 51.1257 & 57.5967 & 53.2340 & \cellcolor{green!25} 60.4816 & \cellcolor{red!25} \textbf{60.6476} \\ 

 	 	 	 	 	 	  & SSIM & 0.9991 & 0.9992 & 0.9992 & \cellcolor{blue!25} 0.9994 & 0.9987 & 0.9989 & 0.9992 & 0.9991 & 0.9982 & \cellcolor{yellow!75} 0.9993 & 0.9973 & \cellcolor{green!25} 0.9996 & \cellcolor{red!25} \textbf{0.9996} \\ 
\bottomrule
\end{tabular}
\end{center}
\end{table*}

%\begin{table}[]
%\caption{Results on MIT-Adobe FiveK dataset~\cite{Adobe}.
%\label{tab:Adobe_1K_test}}
%\begin{center}
%\scriptsize
%\setlength{\tabcolsep}{2pt}
%\begin{tabular}{ccccccccc}
%\toprule
%\multicolumn{2}{c}{} & ZP & MIG & BR & MRC & IPAD & Ours D4 & Ours D16 \\
%\multicolumn{2}{c}{} & & & \cite{BR} & \cite{MRC} & \cite{IPAD}  &  & \\
%\toprule
%\multirow{2}{*}{8-10 bit} & PSNR & 54.7390 & 57.0704 & 56.2218 & \cellcolor{yellow!75} 57.5123 & \cellcolor{blue!25} 57.6648 & \cellcolor{red!25} \textbf{58.8890} & \cellcolor{green!25} 58.8877 \\
%
% 	 	 	 	 	 	  & SSIM & \cellcolor{blue!25} 0.9993 & 0.9992 & 0.9992 & 0.9992 & \cellcolor{yellow!75} 0.9992 & \cellcolor{green!25} 0.9995 & \cellcolor{red!25} \textbf{0.9995} \\
%\midrule
%\multirow{2}{*}{8-12 bit} & PSNR & 53.3305 & 56.4642 & 56.3202 & \cellcolor{blue!25} 58.7387 & \cellcolor{yellow!75} 57.7665 & \cellcolor{green!25} 59.7254 & \cellcolor{red!25} \textbf{59.7783} \\
%
% 	 	 	 	 	 	  & SSIM & 0.9992 & 0.9992 & \cellcolor{yellow!75} 0.9992 & \cellcolor{blue!25} 0.9994 & 0.9991 & \cellcolor{green!25} 0.9995 & \cellcolor{red!25} \textbf{0.9995} \\
%\bottomrule
%\end{tabular}
%\end{center}
%\end{table}

\begin{table*}[]
\caption{Results on MIT-Adobe FiveK dataset~\cite{Adobe}.
\label{tab:Adobe_1K_test}}
\begin{center}
\scriptsize
\setlength{\tabcolsep}{4pt}
\begin{tabular}{cccccccccccc}
\toprule
\multicolumn{2}{c}{} & ZP & MIG & BR & MRC &  IPAD & BE-CNN & BE-CALF & BitNet & Ours D4 & Ours D16 \\
\multicolumn{2}{c}{} & & & \cite{BR} & \cite{MRC} & \cite{IPAD}  & \cite{BE-CNN} & \cite{BE-CALF} & \cite{BitNet} &  & \\
\toprule
\multirow{2}{*}{8-10 bit} & PSNR & 54.7390 & 57.0704 & 56.2218 & \cellcolor{yellow!75} 57.5123 & \cellcolor{blue!25} 57.6648 & 49.8608 & 57.0974 & 57.2695 & \cellcolor{red!25} \textbf{58.8890} & \cellcolor{green!25} 58.8877 \\ 

 	 	 	 	 	 	  & SSIM & \cellcolor{yellow!75} 0.9993 & 0.9992 & 0.9992 & 0.9992 & 0.9992 & 0.9959 & 0.9992 & \cellcolor{blue!25} 0.9993 & \cellcolor{green!25} 0.9995 & \cellcolor{red!25} \textbf{0.9995} \\ 
\midrule
\multirow{2}{*}{8-12 bit} & PSNR & 53.3305 & 56.4642 & 56.3202 & \cellcolor{blue!25} 58.7387 & 57.7665 & 49.8919 & 57.1691 & \cellcolor{yellow!75} 57.8105 & \cellcolor{green!25} 59.7254 & \cellcolor{red!25} \textbf{59.7783} \\ 

 	 	 	 	 	 	  & SSIM & 0.9992 & 0.9992 & 0.9992 & \cellcolor{blue!25} 0.9994 & 0.9991 & 0.9959 & 0.9992 & \cellcolor{yellow!75} 0.9993 & \cellcolor{green!25} 0.9995 & \cellcolor{red!25} \textbf{0.9995} \\ 
\bottomrule
\end{tabular}
\end{center}
\end{table*}

\begin{table*}[]
\caption{6--8 bit recovery.
\label{tab:6_8_bit}}
\begin{center}
\scriptsize
\setlength{\tabcolsep}{4pt}
\begin{tabular}{ccccccccccc}
\toprule
\multicolumn{2}{c}{} & ZP & MIG & BR & MRC & CRR & CA & IPAD & BDEN & Ours D16 \\
\multicolumn{2}{c}{} & & & \cite{BR} & \cite{MRC} & \cite{CRR} & \cite{CA} & \cite{IPAD}  & \cite{BDEN} & \\
\toprule
\multirow{2}{*}{BSD100~\cite{BSD}} & PSNR & 41.1397 & 43.0056 & 42.6987 & \cellcolor{blue!25} 43.9878 & 43.2519 & \cellcolor{yellow!75} 43.3038 & 43.1991 & \cellcolor{green!25} 47.7800$^{\dagger}$ & \cellcolor{red!25} \textbf{48.1065} \\
%48.1065 - 0.9967
 	 	 	 	 	 	  & SSIM & 0.9898 & 0.9896 & 0.9896 & \cellcolor{green!25} 0.9915 & 0.9909 & \cellcolor{yellow!75} 0.9910 & \cellcolor{blue!25} 0.9911 & NR & \cellcolor{red!25} \textbf{0.9967} \\
\midrule
\multirow{2}{*}{Kodak~\cite{Kodak}} & PSNR & 42.5077 & 45.5028 & 45.1540 & \cellcolor{blue!25} 47.1685 & 45.5719 & \cellcolor{yellow!75} 45.7344 & 45.5074 & \cellcolor{green!25} 47.7900$^{\dagger}$ & \cellcolor{red!25} \textbf{47.8255} \\

 	 	 	 	 	 	  & SSIM & \cellcolor{blue!25} 0.9918 & 0.9915 & 0.9915 & \cellcolor{green!25} 0.9929 & 0.9917 & \cellcolor{yellow!75} 0.9918 & 0.9915 & NR & \cellcolor{red!25} \textbf{0.9953} \\
\bottomrule
\end{tabular}
\end{center}
\end{table*}

\begin{table*}
\begin{center}
\caption{Average running times of different algorithms on the Kodak dataset~\cite{Kodak} for 4 to 8 bit recovery.
\label{tab:running_time}}
\scriptsize
\begin{tabular}{c  ccccc cccccccc}
\toprule
Method & ZP & MIG & BR & MRC & CRR & CA & ACDC & IPAD & BE-CNN & BE-CALF & BitNet & Ours D4 & Ours D16 \\
 & & & \cite{BR} & \cite{MRC} & \cite{CRR} & \cite{CA} & \cite{ACDC} & \cite{IPAD} & \cite{BE-CNN} & \cite{BE-CALF} & \cite{BitNet} & & \\
\toprule
Time (s) & 0.006 & 0.002 & 0.004 & 0.641 & 318.86 & 397.51 & 1457.40 & 39.14 & 0.201 & 1.05 & 3.28 & 0.33 & 1.10 \\
\bottomrule
\end{tabular}
\end{center}
\end{table*}

In Table~\ref{tab:8_10_12}, we report results for 8 to 10 bit, and 8 to 12 bit recovery on the Sintel dataset~\cite{Sintel} and TESTIMAGES dataset~\cite{USTHK}. The vast majority of high-bit-depth (HBD) monitors available today are 10 bit, and therefore, 8 to 10 bit reconstruction represents the most common scenario of bit-depth recovery applied to high-bit-depth displays. Modern camera sensors have a dynamic range of 10--12 bits. Thus, 8 to 12 bit recovery is the typical range for photo editing applications. 
%Competing deep neural network (DNN) methods have not provided their results for these quantization levels. Since their pre-trained models/codes are not publicly available, their values are not reported in Table~\ref{tab:8_10_12}. 
It can be seen that our proposed method obtains higher accuracy than competing approaches. Results on MIT-Adobe FiveK dataset~\cite{Adobe} for the same quantization levels are provided in Table~\ref{tab:Adobe_1K_test}. Comparisons against CRR~\cite{CRR}, CA~\cite{CA}, and ACDC~\cite{ACDC} have been omitted since there are 1000 images in the test set, and the running times (see Table~\ref{tab:running_time}) of these methods are very high. 
%Moreover, from Tables 1 to 5 of the main paper, and Table~\ref{tab:UST_HK_800} of the supplementary material, it can be observed that IPAD~\cite{IPAD}, on average, performs the best among the classical methods, and the results of this method have been included in Table~\ref{tab:Adobe_1K_test}. 
%Our method outperforms IPAD~\cite{IPAD} and other competitors, such as MRC~\cite{MRC}, by a sound margin.
Our method again outperforms competitors by a sound margin.

BDEN~\cite{BDEN} has reported scores on the Kodak dataset~\cite{Kodak} and the BSD100 dataset for 6 to 8 bit recovery. The BSD100 dataset consists of 100 images selected from the Berkeley Segmentation dataset~\cite{BSD}. A demo version of the code for the deep learning method in BDEN has been publicly released by the authors~\cite{BDEN}. However, it works only on grayscale images, and the model has been trained using only the first 100 images from the DIV2K~\cite{DIV2K} dataset, while their actual model in the paper is trained on color images using all 800 images from this dataset. In consideration of these limitations, for fair comparison, we directly copy results reported in their paper instead of running their demo code. The results are presented in Table~\ref{tab:6_8_bit}. We would like to note that our bitplane-wise recovery strategy offers the most advantage over single-shot methods for larger expansions of the bit depth e.g., 8--12 bit, 6--12 bit etc. In the present case, the expansion is only 2 bits (from 6 to 8), and hence, our PSNR values are only slightly higher than BDEN.

\section{Additional qualitative results}
\label{sec:qualitative}
\begin{figure*}[!t]
\begin{center}
\setlength{\tabcolsep}{2pt}
\begin{tabular}{c}
\includegraphics[width=0.29\linewidth]{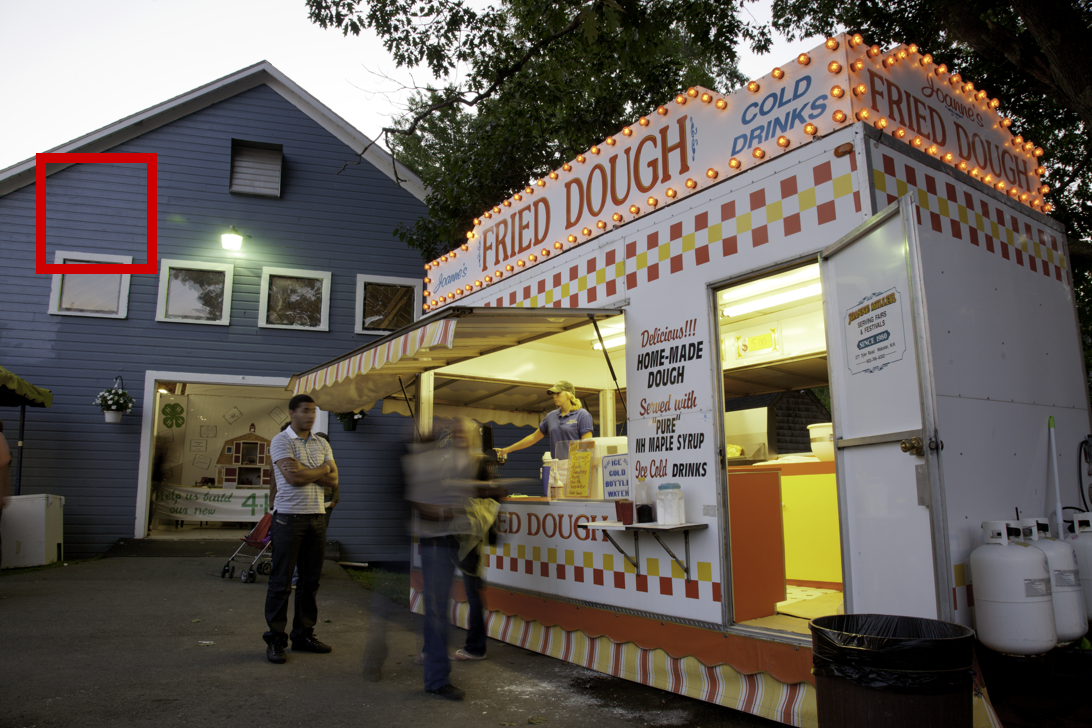}
\end{tabular}
\\ Ground truth (GT)
\begin{tabular}{ccccc}
\includegraphics[width=0.18\linewidth]{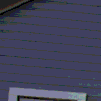}&
\includegraphics[width=0.18\linewidth]{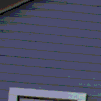}&
\includegraphics[width=0.18\linewidth]{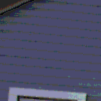}&
\includegraphics[width=0.18\linewidth]{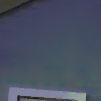}&
\includegraphics[width=0.18\linewidth]{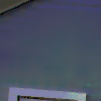}\\
ZP & BR~\cite{BR} & MRC~\cite{MRC} & CRR~\cite{CRR} & CA~\cite{CA} \\

\includegraphics[width=0.18\linewidth]{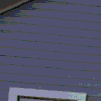}&
\includegraphics[width=0.18\linewidth]{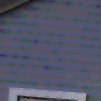}&
\includegraphics[width=0.18\linewidth]{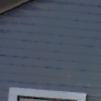}&
\includegraphics[width=0.18\linewidth]{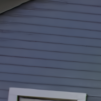}&
\includegraphics[width=0.18\linewidth]{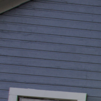}\\
ACDC~\cite{ACDC} & IPAD~\cite{IPAD} & BitNet~\cite{BitNet} & Ours D16 & GT\\
\end{tabular}

\begin{tabular}{c}
\includegraphics[width=0.29\linewidth]{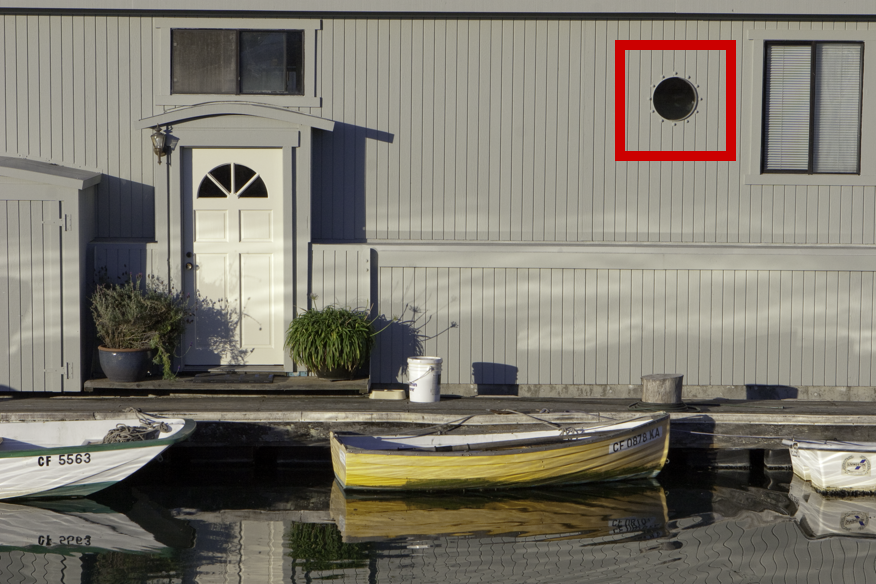}
\end{tabular}
\\ Ground truth (GT)
\begin{tabular}{ccccc}
\includegraphics[width=0.18\linewidth]{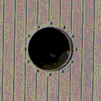}&
\includegraphics[width=0.18\linewidth]{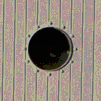}&
\includegraphics[width=0.18\linewidth]{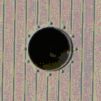}&
\includegraphics[width=0.18\linewidth]{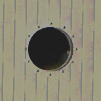}&
\includegraphics[width=0.18\linewidth]{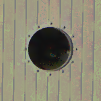}\\
ZP & BR~\cite{BR} & MRC~\cite{MRC} & CRR~\cite{CRR} & CA~\cite{CA}\\

\includegraphics[width=0.18\linewidth]{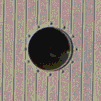}&
\includegraphics[width=0.18\linewidth]{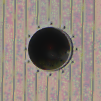}&
\includegraphics[width=0.18\linewidth]{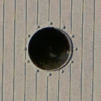}&
\includegraphics[width=0.18\linewidth]{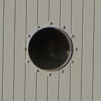}&
\includegraphics[width=0.18\linewidth]{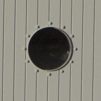}\\
ACDC~\cite{ACDC} & IPAD~\cite{IPAD} & BitNet~\cite{BitNet} & Ours D16 & GT\\
\end{tabular}

\end{center}
\caption{Qualitative comparisons on the MIT-Adobe FiveK dataset~\cite{Adobe} for 3 to 8 bit recovery.
\label{fig:adobe}}
\end{figure*}

\begin{figure*}[!t]
\begin{center}
\setlength{\tabcolsep}{2pt}
\begin{tabular}{c}
\includegraphics[width=0.44\linewidth]{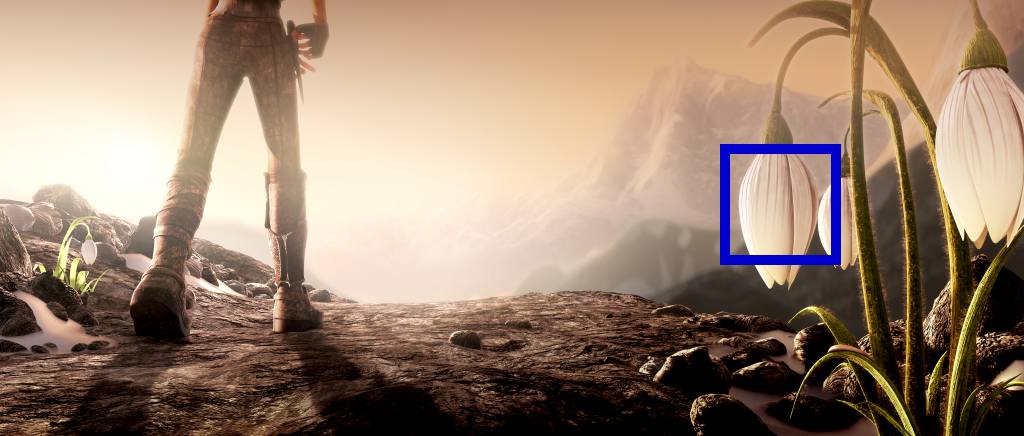}
\end{tabular}
\\ Ground truth (GT)
\begin{tabular}{ccccc}
\includegraphics[width=0.18\linewidth]{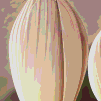}&
\includegraphics[width=0.18\linewidth]{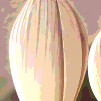}&
\includegraphics[width=0.18\linewidth]{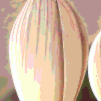}&
\includegraphics[width=0.18\linewidth]{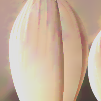} &
\includegraphics[width=0.18\linewidth]{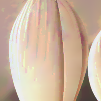}\\
ZP & BR~\cite{BR} & MRC~\cite{MRC} & CRR~\cite{CRR} & CA~\cite{CA} \\

\includegraphics[width=0.18\linewidth]{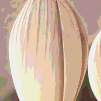}&
\includegraphics[width=0.18\linewidth]{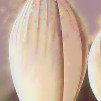}&
\includegraphics[width=0.18\linewidth]{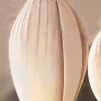}&
\includegraphics[width=0.18\linewidth]{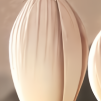}&
\includegraphics[width=0.18\linewidth]{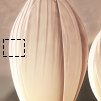}\\
ACDC~\cite{ACDC} & IPAD~\cite{IPAD} & BitNet~\cite{BitNet} & Ours D16 & GT \\
\end{tabular}

\begin{tabular}{c}
\includegraphics[width=0.44\linewidth]{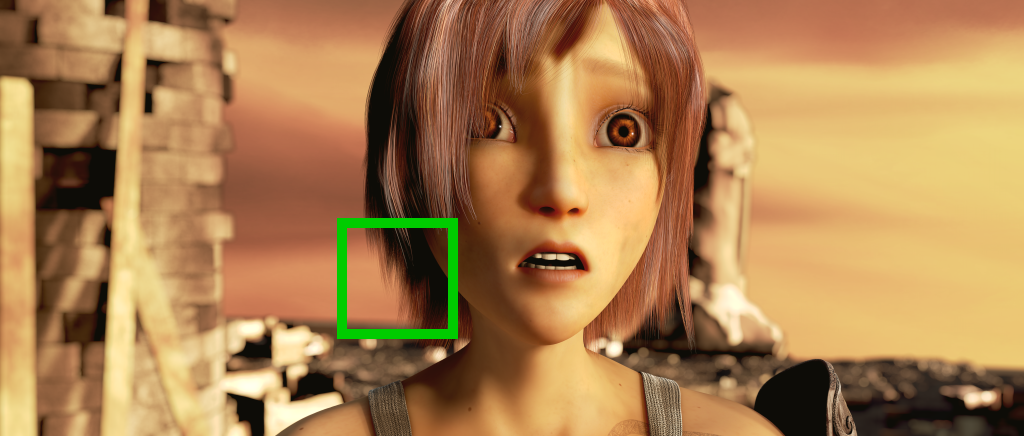}
\end{tabular}
\\ Ground truth (GT)
\begin{tabular}{ccccc}
\includegraphics[width=0.18\linewidth]{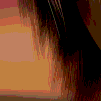}&
\includegraphics[width=0.18\linewidth]{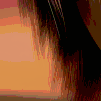}&
\includegraphics[width=0.18\linewidth]{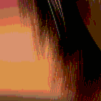}&
\includegraphics[width=0.18\linewidth]{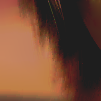}&
\includegraphics[width=0.18\linewidth]{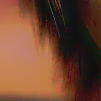}\\
ZP & BR~\cite{BR} & MRC~\cite{MRC} & CRR~\cite{CRR} & CA~\cite{CA}\\

\includegraphics[width=0.18\linewidth]{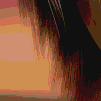}&
\includegraphics[width=0.18\linewidth]{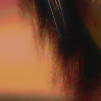}&
\includegraphics[width=0.18\linewidth]{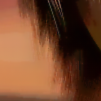}&
\includegraphics[width=0.18\linewidth]{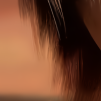}&
\includegraphics[width=0.18\linewidth]{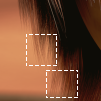}\\
ACDC~\cite{ACDC} & IPAD~\cite{IPAD} & BitNet~\cite{BitNet} & Ours D16 & GT \\
\end{tabular}

\end{center}
\caption{Qualitative comparisons on the Sintel dataset~\cite{Sintel} for 3 to 8 bit recovery.
\label{fig:Sintel}}
\end{figure*}

\begin{figure*}[!t]
\begin{center}
\setlength{\tabcolsep}{2pt}
\begin{tabular}{c}
\includegraphics[width=0.2\linewidth]{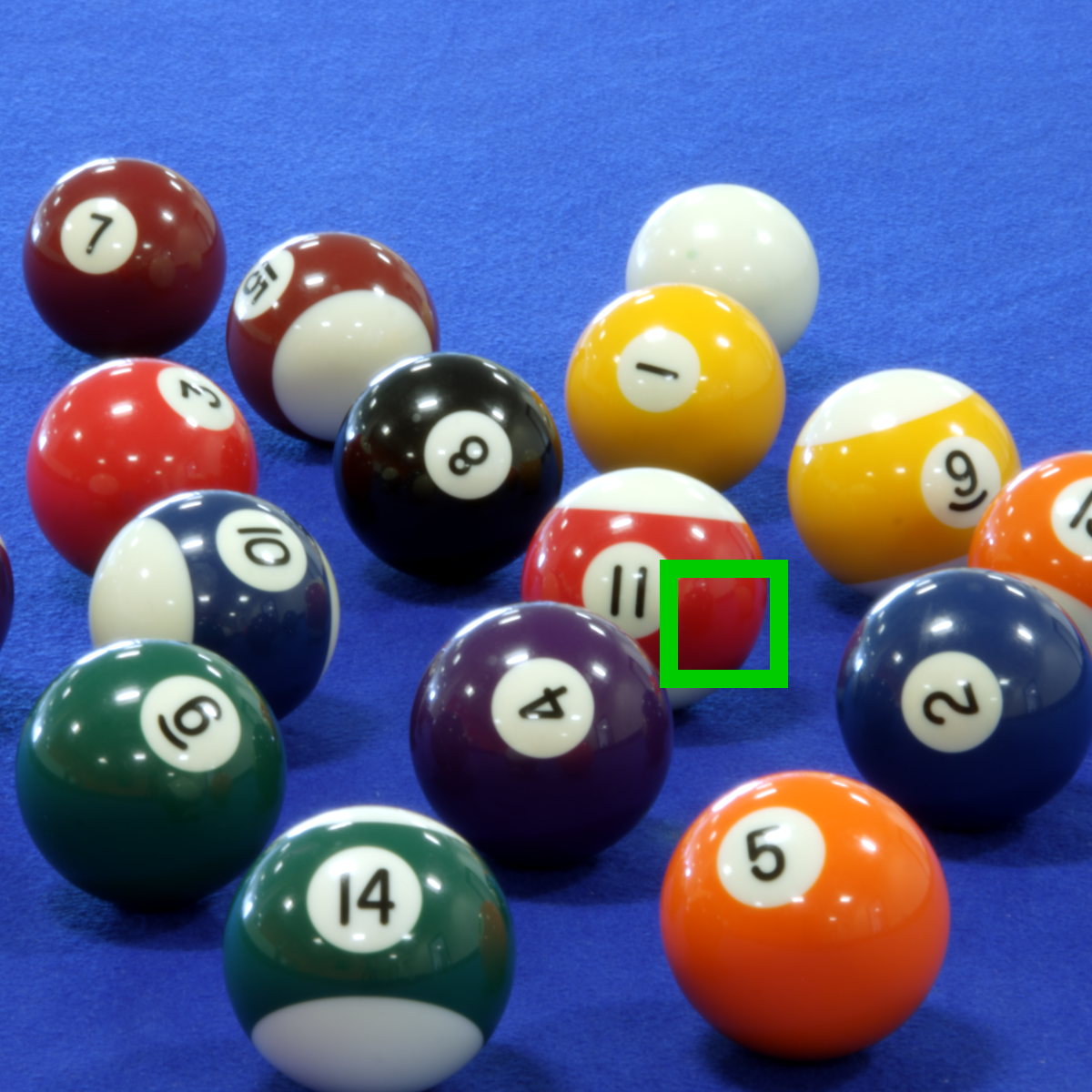}
\end{tabular}
\\ Ground truth (GT)
\begin{tabular}{ccccc}
\includegraphics[width=0.18\linewidth]{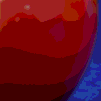}&
\includegraphics[width=0.18\linewidth]{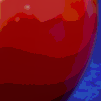}&
\includegraphics[width=0.18\linewidth]{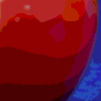}&
\includegraphics[width=0.18\linewidth]{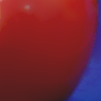} &
\includegraphics[width=0.18\linewidth]{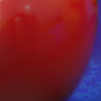}\\
ZP & BR~\cite{BR} & MRC~\cite{MRC} & CRR~\cite{CRR} & CA~\cite{CA} \\

\includegraphics[width=0.18\linewidth]{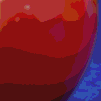}&
\includegraphics[width=0.18\linewidth]{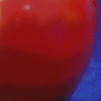}&
\includegraphics[width=0.18\linewidth]{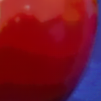}&
\includegraphics[width=0.18\linewidth]{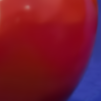}&
\includegraphics[width=0.18\linewidth]{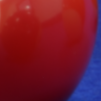}\\
ACDC~\cite{ACDC} & IPAD~\cite{IPAD} & BitNet~\cite{BitNet} & Ours D16 & GT \\
\end{tabular}

\begin{tabular}{c}
\includegraphics[width=0.2\linewidth]{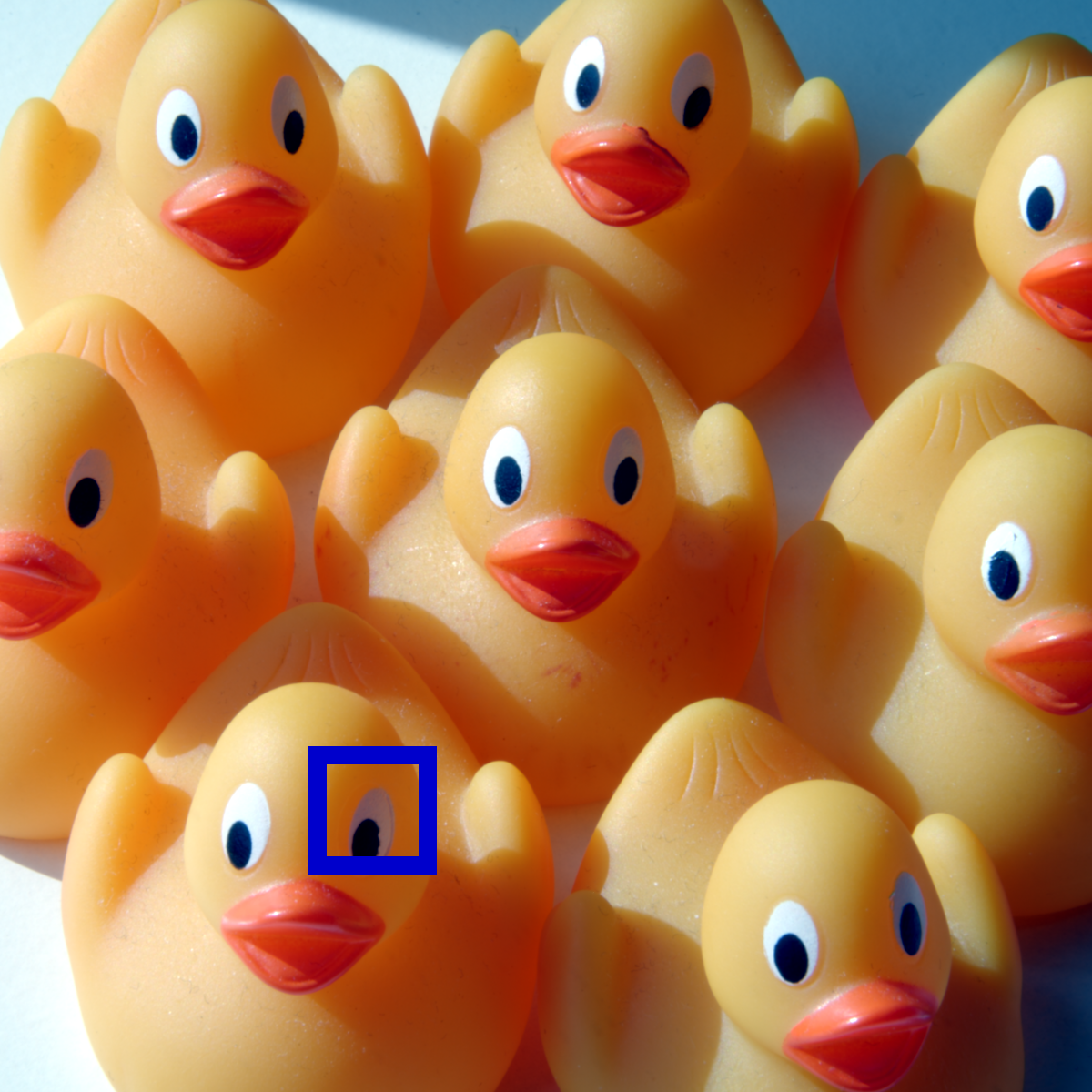}
\end{tabular}
\\ Ground truth (GT)
\begin{tabular}{ccccc}
\includegraphics[width=0.18\linewidth]{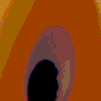}&
\includegraphics[width=0.18\linewidth]{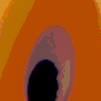}&
\includegraphics[width=0.18\linewidth]{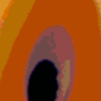}&
\includegraphics[width=0.18\linewidth]{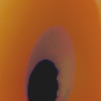}&
\includegraphics[width=0.18\linewidth]{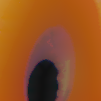}\\
ZP & BR~\cite{BR} & MRC~\cite{MRC} & CRR~\cite{CRR} & CA~\cite{CA}\\

\includegraphics[width=0.18\linewidth]{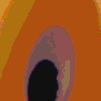}&
\includegraphics[width=0.18\linewidth]{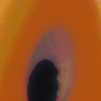}&
\includegraphics[width=0.18\linewidth]{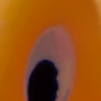}&
\includegraphics[width=0.18\linewidth]{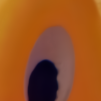}&
\includegraphics[width=0.18\linewidth]{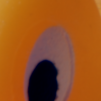}\\
ACDC~\cite{ACDC} & IPAD~\cite{IPAD} & BitNet~\cite{BitNet} & Ours D16 & GT \\
\end{tabular}

\end{center}
\caption{Qualitative comparisons on the TESTIMAGES 1200 dataset~\cite{USTHK} for 3 to 8 bit recovery.
\label{fig:1200}}
\end{figure*}

\begin{figure*}[!t]
\begin{center}
\setlength{\tabcolsep}{2pt}
\begin{tabular}{c}
\includegraphics[width=0.3\linewidth]{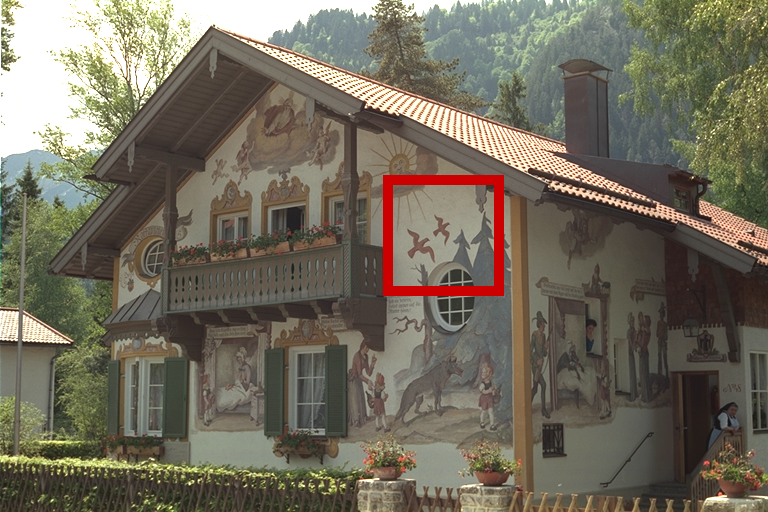}
\end{tabular}
\\ Ground truth (GT)
\begin{tabular}{ccccc}
\includegraphics[width=0.18\linewidth]{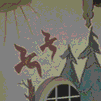}&
\includegraphics[width=0.18\linewidth]{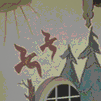}&
\includegraphics[width=0.18\linewidth]{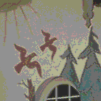}&
\includegraphics[width=0.18\linewidth]{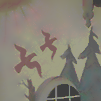} &
\includegraphics[width=0.18\linewidth]{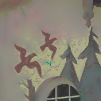}\\
ZP & BR~\cite{BR} & MRC~\cite{MRC} & CRR~\cite{CRR} & CA~\cite{CA} \\

\includegraphics[width=0.18\linewidth]{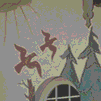}&
\includegraphics[width=0.18\linewidth]{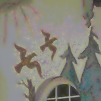}&
\includegraphics[width=0.18\linewidth]{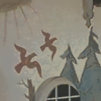}&
\includegraphics[width=0.18\linewidth]{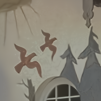}&
\includegraphics[width=0.18\linewidth]{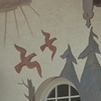}\\
ACDC~\cite{ACDC} & IPAD~\cite{IPAD} & BitNet~\cite{BitNet} & Ours D16 & GT \\
\end{tabular}

\begin{tabular}{c}
\includegraphics[width=0.3\linewidth]{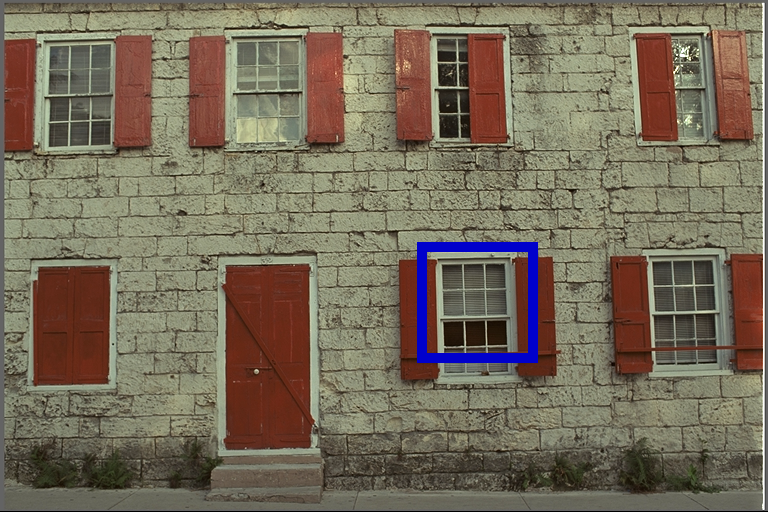}
\end{tabular}
\\ Ground truth (GT)
\begin{tabular}{ccccc}
\includegraphics[width=0.18\linewidth]{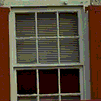}&
\includegraphics[width=0.18\linewidth]{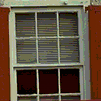}&
\includegraphics[width=0.18\linewidth]{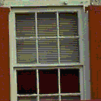}&
\includegraphics[width=0.18\linewidth]{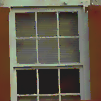}&
\includegraphics[width=0.18\linewidth]{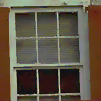}\\
ZP & BR~\cite{BR} & MRC~\cite{MRC} & CRR~\cite{CRR} & CA~\cite{CA}\\

\includegraphics[width=0.18\linewidth]{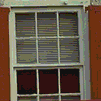}&
\includegraphics[width=0.18\linewidth]{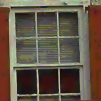}&
\includegraphics[width=0.18\linewidth]{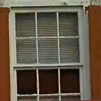}&
\includegraphics[width=0.18\linewidth]{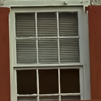}&
\includegraphics[width=0.18\linewidth]{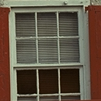}\\
ACDC~\cite{ACDC} & IPAD~\cite{IPAD} & BitNet~\cite{BitNet} & Ours D16 & GT \\
\end{tabular}

\end{center}
\caption{Qualitative comparisons on the Kodak dataset~\cite{Kodak} for 3 to 8 bit recovery.
\label{fig:Kodak}}
\end{figure*}

\begin{figure*}[!t]
\begin{center}
\setlength{\tabcolsep}{2pt}
\begin{tabular}{c}
\includegraphics[width=0.35\linewidth]{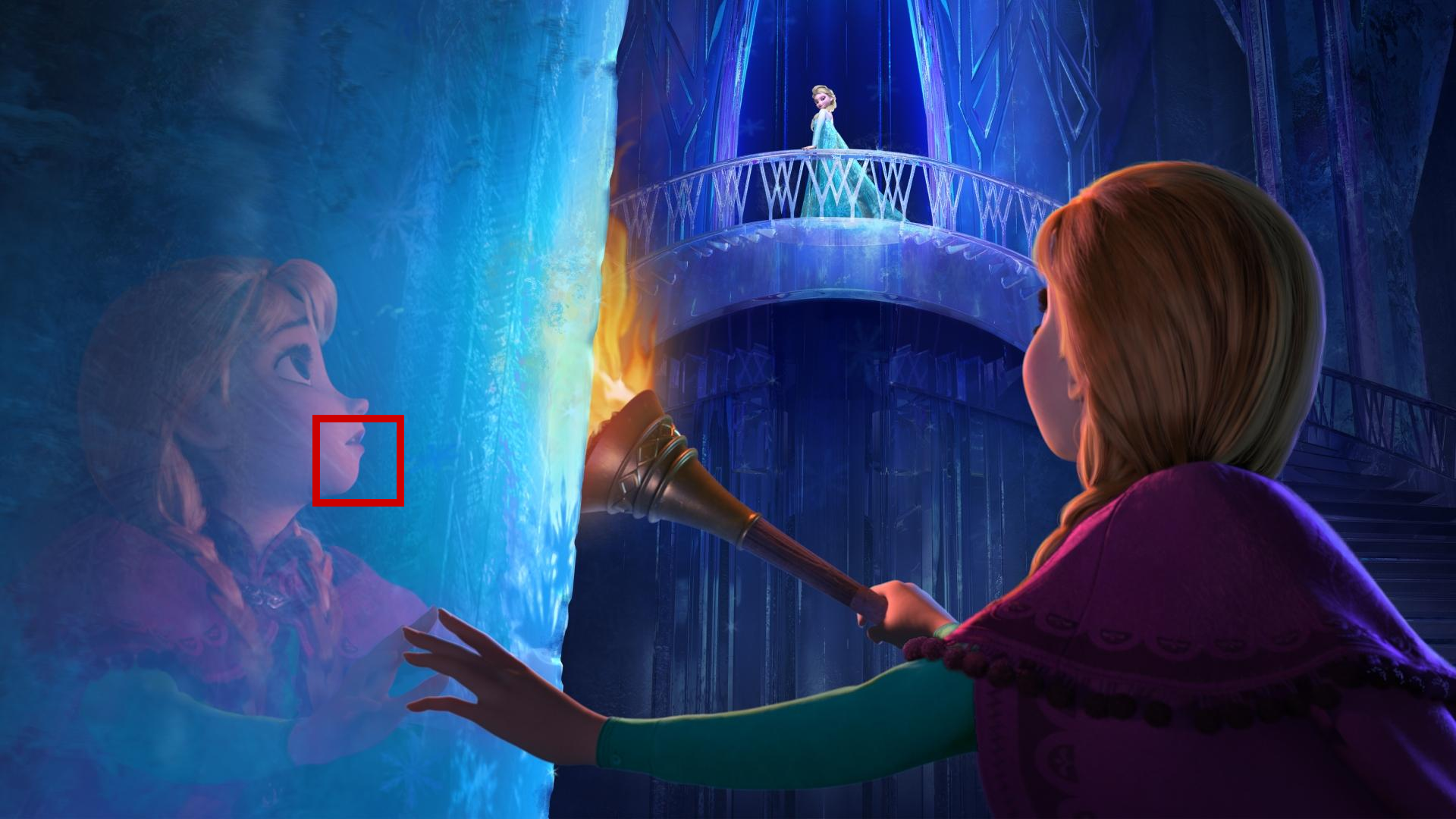}
\end{tabular}
\\ Ground truth (GT)
\begin{tabular}{ccccc}
\includegraphics[width=0.18\linewidth]{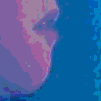}&
\includegraphics[width=0.18\linewidth]{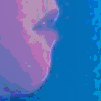}&
\includegraphics[width=0.18\linewidth]{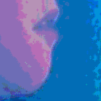}&
\includegraphics[width=0.18\linewidth]{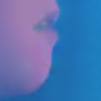} &
\includegraphics[width=0.18\linewidth]{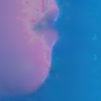}\\
ZP & BR~\cite{BR} & MRC~\cite{MRC} & CRR~\cite{CRR} & CA~\cite{CA} \\

\includegraphics[width=0.18\linewidth]{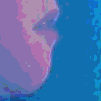}&
\includegraphics[width=0.18\linewidth]{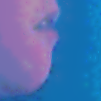}&
\includegraphics[width=0.18\linewidth]{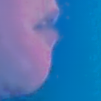}&
\includegraphics[width=0.18\linewidth]{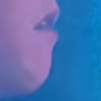}&
\includegraphics[width=0.18\linewidth]{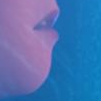}\\
ACDC~\cite{ACDC} & IPAD~\cite{IPAD} & BitNet~\cite{BitNet} & Ours D16 & GT \\
\end{tabular}

\begin{tabular}{c}
\includegraphics[width=0.35\linewidth]{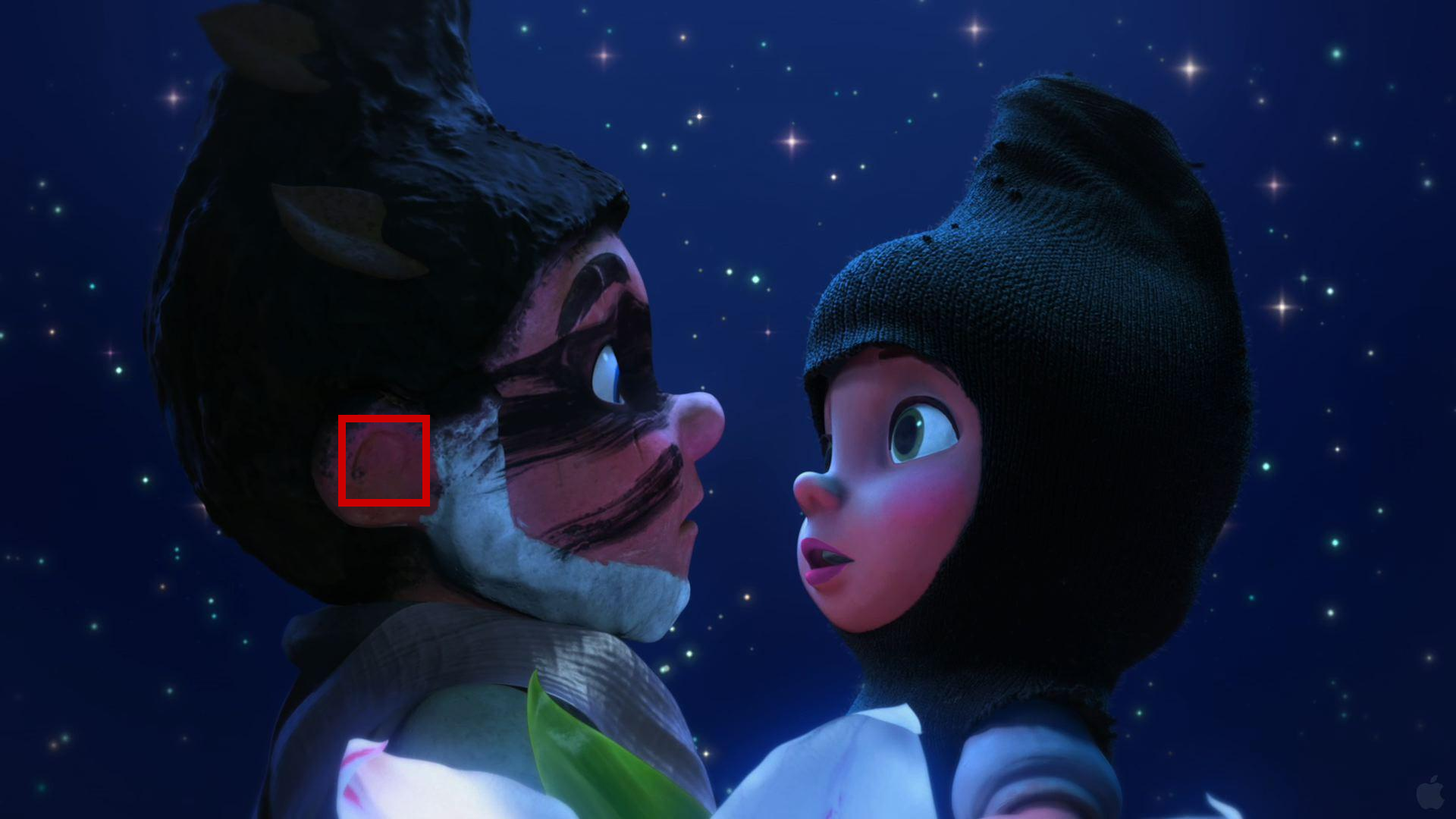}
\end{tabular}
\\ Ground truth (GT)
\begin{tabular}{ccccc}
\includegraphics[width=0.18\linewidth]{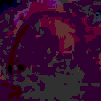}&
\includegraphics[width=0.18\linewidth]{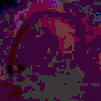}&
\includegraphics[width=0.18\linewidth]{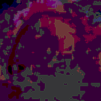}&
\includegraphics[width=0.18\linewidth]{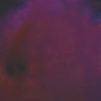}&
\includegraphics[width=0.18\linewidth]{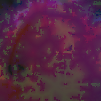}\\
ZP & BR~\cite{BR} & MRC~\cite{MRC} & CRR~\cite{CRR} & CA~\cite{CA}\\

\includegraphics[width=0.18\linewidth]{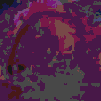}&
\includegraphics[width=0.18\linewidth]{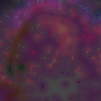}&
\includegraphics[width=0.18\linewidth]{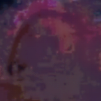}&
\includegraphics[width=0.18\linewidth]{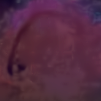}&
\includegraphics[width=0.18\linewidth]{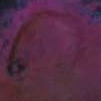}\\
ACDC~\cite{ACDC} & IPAD~\cite{IPAD} & BitNet~\cite{BitNet} & Ours D16 & GT \\
\end{tabular}

\end{center}
\caption{Qualitative comparisons on the ESPL v2 dataset~\cite{ESPL} for 3 to 8 bit recovery.
\label{fig:ESPL}}
\end{figure*}

Qualitative comparisons for 3 to 8 bit recovery on MIT-Adobe FiveK~\cite{Adobe}, Sintel~\cite{Sintel}, TESTIMAGES 1200~\cite{USTHK}, Kodak~\cite{Kodak}, and ESPL v2~\cite{ESPL} datasets are provided in Figs. \ref{fig:adobe}, \ref{fig:Sintel}, \ref{fig:1200}, \ref{fig:Kodak} and \ref{fig:ESPL}, respectively. Zoomed-in patches are presented for better visualization. Representative structures including smooth color transitions, edges, and fine textures are shown. 
%As already mentioned in the main paper, our qualitative comparisons are limited to the eight classical methods since the pre-trained models/codes of competing DNN methods are unavailable.
%Furthermore, regions of interest are marked with a circle on the zoomed-in ground truth patch in Figs. \ref{fig:qualitative3}, and \ref{fig:qualitative2}.

In the first example of Fig. \ref{fig:adobe}, it can be clearly observed that the horizontal lines are best recovered by our method. In the second example, competing approaches cannot reproduce the correct color tones, and have artifacts inside the circle, while our result is closest to the ground truth. In Fig. \ref{fig:Sintel} from the Sintel dataset, the thin petal structures and the fine details in the hair (denoted by dotted rectangles on the zoomed-in ground truth patches) are best reconstructed by our method as compared to competitors. Fig. \ref{fig:1200} shows examples of smooth color transitions, and it can be seen that other algorithms introduce false edges (first image) or fail to detect the boundary of the eye (second image). In Fig. \ref{fig:Kodak}, comparison techniques fail to recover the fine lines on the top left in the first example, while artifacts can clearly be observed in the window blinds in the second example. Our approach produces an accurate reconstruction in both cases. In the first animated example from the ESPL v2 dataset in Fig. \ref{fig:ESPL}, it can be seen that most competing methods, including BitNet~\cite{BitNet}, produce artifacts along the edge of the face, particularly in the bottom left. Our result more closely matches the ground truth. The second image is a challenging example where we yet again produce compelling results, whereas comparison methods fail to recover the image structure.

\section{Running time}
\label{sec:time}
In Table \ref{tab:running_time}, we report the average running times of eight classical methods and three DNN methods on the Kodak dataset~\cite{Kodak} which contains images of size $768 \times 512$ pixels. The 4 to 8 bit recovery scenario is tested. As already mentioned in the main paper, we used the codes made publicly available by the authors of~\cite{IPAD} to compare against the classical methods. The implementations are CPU-based. In the last two columns of Table \ref{tab:running_time}, the average running times of our D4 and D16 models to process four bitplanes are reported. Our method is implemented on GPU. Although ZP, MIG, and BR~\cite{BR} are extremely fast, their performance is poor. CRR~\cite{CRR}, CA~\cite{CA}, and ACDC~\cite{ACDC} have high computational complexity, and their outputs are inferior to ours both qualitatively and quantitatively.
%Again, IPAD~\cite{IPAD} performs the best among the classical methods.
Although the difference in the testing environment (CPU versus GPU) biases a direct comparison of running times in our favour, it is still noteworthy that our proposed method is much faster than IPAD, which is the best performing classical method. More importantly, our method produces more accurate results than IPAD. The three DNN methods BE-CNN~\cite{BE-CNN}, BE-CALF~\cite{BE-CALF} and BitNet~\cite{BitNet}, as well as our algorithm were tested on an Nvidia Tesla V100 GPU with 32 GB of RAM. While BE-CNN is faster than our method, its results are poor. In terms of accuracy, our two closest competitors are BE-CALF and BitNet. Our D16 model has nearly the same running time as BE-CALF, while D4 is faster. Both our models are significantly faster than BitNet.

\section{Photo editing application}
\begin{figure*}
\begin{center}
\includegraphics[width=0.65\linewidth]{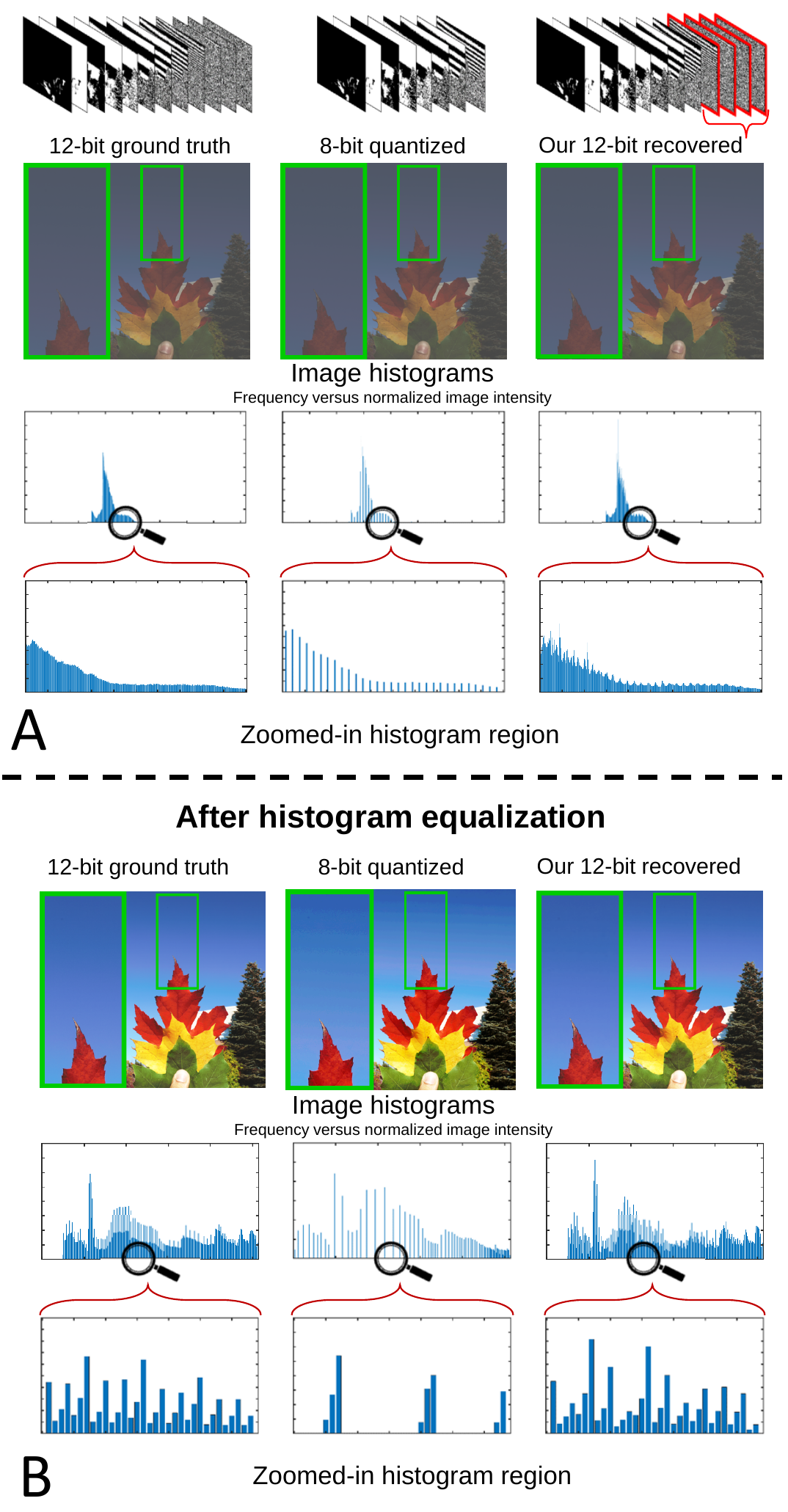}
\caption{(A) Shows a ground truth 12-bit HBD image, its 8-bit quantized LBD version, and our recovered 12-bit HBD result. The image has low contrast as shown by its image histogram.  The green box denotes a zoomed-in region of the image. (B) Shows the same three images with histogram equalization applied.  Due to its low bit depth, the 8-bit image has gaps in its histogram and exhibits visual banding. Our recovered 12-bit HBD image has a similar visual appearance and histogram profile as the ground truth 12-bit HBD image.
\label{fig:editing_1}}
\end{center}
\end{figure*}

The ability to edit an image post-capture can be adversely affected by quantization to 8 bits. Fig. \ref{fig:editing_1}-(A) shows a common example of a captured 8-bit sRGB image having low contrast. Applying a simple histogram equalization operation on this 8-bit image to enhance its contrast produces the undesired ``stair-step'' effect shown in Fig. \ref{fig:editing_1}-(B).  This artifact is a direct consequence of the quantization process whereby tonal information was lost.  Fig. \ref{fig:editing_1}-(B) shows the results of histogram equalization when applied to the original unquantized 12-bit image (which is usually unavailable) and our recovered 12-bit result. Due to the additional tonal values in the 12-bit histograms, the banding artifacts are not present in these outputs. This example highlights the importance of bit-depth recovery for photo editing.
%In Fig. 1 of the main paper, we had highlighted the importance of bit-depth recovery for post-capture photo editing. An additional example is provided in Fig.~\ref{fig:editing_1}. As seen from the results, applying our bit-depth recovery algorithm on the 8-bit image as a pre-processing step improves the ability to enhance contrast without introducing banding artifacts. For comparison, the result of IPAD is also provided. It can be seen from the zoomed-in patch that their output still exhibits some visual banding.

\label{sec:editing}

\end{document}